\documentclass[%
 reprint,
 aps,pra,
 amsmath,amssymb,
 groupedaddress,
 longbibliography,
]{revtex4-1}

\usepackage{graphicx}% Include figure files
\usepackage{hyperref}

\begin{document}

\title{Enhanced high-order harmonic generation in donor-doped band-gap materials}

\author{Chuan Yu}
% \email{chuan@phys.au.dk}
\author{Kenneth K. Hansen}
% \email{klochmann@phys.au.dk}
\author{Lars Bojer Madsen}
% \email{bojer@phys.au.dk}
\affiliation{Department of Physics and Astronomy, Aarhus University, DK-8000 Aarhus C, Denmark}

\date{\today}

\begin{abstract}

We find that a donor-doped band-gap material can enhance the overall high-order harmonic generation (HHG) efficiency by several orders of magnitude, compared with undoped and acceptor-doped materials.
This significant enhancement, predicted by time-dependent density functional theory simulations, originates from the highest occupied impurity state which has an isolated energy located within the band gap.
The impurity-state HHG is rationalized by a three-step model, taking into account that the impurity-state electron tunnels into the conduction band and then moves according to its band structure until recombination.
In addition to the improvement of the HHG efficiency, the donor-type doping results in a harmonic cutoff different from that in the undoped and acceptor-doped cases, explained by semiclassical analysis for the impurity-state HHG.

\end{abstract}

\maketitle

\section{\label{sec:intro}Introduction}

High-order harmonic generation (HHG) in gases \cite{hhg1992prl,hhg1993prl} is not only one of the fundamental strong-field phenomena in laser-matter interactions. It is also a powerful technique to produce sub-femtosecond laser pulses, providing the opportunity to explore ultrafast dynamics in matter on femto- and attosecond timescales \cite{atto2009rmp}.
Recently, HHG in solids was demonstrated \cite{ghimire2011natphys,schubert2014natphoton,luu2015nature,vampa2015nature,hohenleutner2015nature,ndabashimiye2016nature} with potential applications for novel VUV and XUV light sources and for probing ultrafast dynamics in condensed-matter systems \cite{attosolid2018rmp}.
Compared with gas-phase systems, solids can possibly produce HHG more efficiently due to their periodic structure and high density.
Also, laser-induced processes in bulk and nanostructured materials attract theoretical interests in this new research area where strong-field laser physics meets condensed matter.
It has been demonstrated that some strong-field concepts, such as the three-step model for HHG \cite{threestep1993prl}, can be generalized to describe laser-solid interactions when the band structure is taken into account \cite{vampa2014prl,vampa2017jpb}.
Although the understanding of HHG in solids is rapidly expanding \cite{higuchi2014prl,vampa2014prl,vampa2017jpb,vampa2015prb,cmcdo2015pra,hawkins2015pra,wmx2015pra,wmx2016pra,bxb2016pra,bxb2018pra1,bxb2018pra2,jin2018jpb,luu2016prb,foeldi2017prb,osika2017prx,rubio2017prl,kkh2017pra,kkh2018pra,kkh2018prl,murakami2018prl,silva2018natphoton,luu2018natcommun},
many open questions remain to be explored.

For applications of HHG in solids as a coherent VUV or XUV source, a key question is how to control the harmonic yield.
A recent experiment has demonstrated enhanced HHG emission in tailored semiconductors \cite{sivis2017science}.
Theoretical studies have proposed possibilities to enhance HHG in solids by quantum confinement \cite{cmcdo2017prl2}, inhomogeneous fields \cite{bxb2016pra2}, or substitutional doping \cite{huang2017pra}.
Indeed, impurities typically influence the physical properties of a solid, allowing one to control processes in the target material for various applications (see, e.g., the recent works \cite{impurity2016natcommun,impurity2018prl}).
Doping-induced impurities are therefore expected to have impact on HHG in solids.
The specific influence of doping-induced impurities, however, still requires further exploration, even in the case of substitutional doping \footnote{The periodic arrangement of dopants at a very high impurity concentration considered in the single-active-electron model calculations of Ref.~\cite{huang2017pra} is atypical and leads to a substantial change  of the system.}.
Here, to elucidate effects of substitutional doping on HHG in solids, we consider a model of undoped and doped band-gap materials interacting with a mid-infrared laser pulse, use time-dependent density functional theory (TDDFT) \cite{tddft1984prl,ullrich2011book} to perform self-consistent calculations, and provide a semiclassical analysis for the impurity-state HHG cutoff.

This paper is organized as follows.
In Sec.~\ref{sec:model}, we describe the theoretical model and methods used in this work.
The results of our theoretical calculations are presented and discussed in Sec.~\ref{sec:result}.
Finally we conclude with a brief summary in Sec.~\ref{sec:concl}.
Atomic units (a.u.) are used throughout unless stated otherwise.

\section{\label{sec:model}Theoretical Model and Methods}

Our model employs a finite system so large that it behaves like a solid \cite{kkh2017pra,kkh2018pra,kkh2018prl}.
We consider a linear chain of $N$ nuclei with a separation $a$ and located at $x_{j}=[j-(N+1)/2]a$, ($j=1,\cdots,N$).
The ionic potential reads $v_{\text{ion}}^{}(x)=-\sum_{j=1}^{N}Z_{j}[(x-x_{j})^{2}+\epsilon]^{-1/2}$,
% \begin{equation}
% v_{\text{ion}}^{}(x) = -\sum_{j=1}^{N} \frac{Z_{j}}{\sqrt{(x-x_{j})^{2}+\epsilon}}, \label{eq:ionpot}
% \end{equation}
where $Z_{j}$ is the nuclear charge of the $j$-th ion and $\epsilon$ is a softening parameter which smoothens the Coulomb singularity.
We set $\epsilon=2.25$ and $a=7$ throughout, and use $Z_{j}=4$ ($j=1,\cdots,N$) to qualitatively model an undoped band-gap material.
For a convenient description of substitutional doping, we choose an odd number of nuclei ($N=2M-1=101$ in this work) and introduce an impurity in the center by choosing a different nuclear charge of the $M$-th ion.
As we will see below, such a doping rate of $\sim1\%$ does not change the band structures significantly, but introduces new states that are energetically isolated.
Also, our discussion of doping effects is insensitive to the model size, once a sufficiently large number of nuclei ($\gtrsim80$) is considered (see the Appendix).
In our simulations, two doping cases are considered: $Z_{M}=2$ for modeling a double acceptor and $Z_{M}=6$ for modeling a double donor \cite{impurity1978rmp}.
All the considered systems are charge and spin neutral.
Thus the number of electrons with opposite spin is $N_{\downarrow}=N_{\uparrow}=2N$ for the undoped case, and $N_{\downarrow}=N_{\uparrow}=2N\pm1$ for the systems with a doped center ($Z_{M}=4\pm2$).
We treat the field-free electronic states for these systems with density functional theory (DFT).
In the Kohn-Sham (KS) scheme, we find a set of KS orbitals determined by
\begin{equation}
\left\{-\frac{1}{2}\frac{\partial^{2}}{\partial x^{2}} + v_{\text{KS}}^{}[\{n_{\sigma}^{}\}](x)\right\}\varphi_{j,\sigma}^{}(x) = \varepsilon_{j,\sigma}^{}\varphi_{j,\sigma}^{}(x), \label{eq:stat_kseq}
\end{equation}
with the static KS potential $v_{\text{KS}}^{}[\{n_{\sigma}^{}\}](x)=v_{\text{ion}}^{}(x)+v_{\text{H}}^{}[n](x)+v_{\text{xc}}^{}[\{n_{\sigma}^{}\}](x)$.
% \begin{equation}
% v_{\text{KS}}^{}[\{n_{\sigma}^{}\}](x) = v_{\text{ion}}^{}(x) + v_{\text{H}}^{}[n](x) + v_{\text{xc}}^{}[\{n_{\sigma}^{}\}](x). \label{eq:stat_kspot}
% \end{equation}
The Hartree potential reads $v_{\text{H}}^{}[n](x)=\int{dx'}n(x')[(x-x')^{2}+\epsilon]^{-1/2}$,
% \begin{equation}
% v_{\text{H}}^{}[n](x) = \int dx' \frac{n(x')}{\sqrt{(x-x')^{2}+\epsilon}}, \label{eq:harpot}
% \end{equation}
and the exchange-correlation potential is treated in a local spin-density approximation $v_{\text{xc}}^{}[{n_{\sigma}}](x)\simeq v_{\text{x}}^{}[{n_{\sigma}^{}}](x)=-[6n_{\sigma}(x)/\pi]^{1/3}$.
% \begin{equation}
% v_{\text{xc}}^{}[{n_{\sigma}}](x) \simeq v_{\text{x}}^{}[{n_{\sigma}^{}}](x) = -\left[\frac{6}{\pi}n_{\sigma}(x)\right]^{1/3}. \label{eq:xcpot}
% \end{equation}
The spin densities are $n_{\sigma}^{}(x)=\sum_{j=1}^{N_{\sigma}}|\varphi_{j,\sigma}^{}(x)|^{2}$ for spin $\sigma={\downarrow,\uparrow}$, and the total density is $n(x)=\sum_{\sigma=\downarrow,\uparrow}n_{\sigma}^{}(x)$.
% \begin{equation}
% n_{\sigma}^{}(x) = \sum_{j=1}^{N_{\sigma}}|\varphi_{j,\sigma}^{}(x)|^{2}, \quad n(x) = \sum_{\sigma=\downarrow,\uparrow} n_{\sigma}^{}(x).
% \end{equation}
% The LSD correlation part is neglected, as it should not affect the qualitative findings in our simulations.

For the driving laser pulse linearly polarized along the $x$-axis, we use the vector potential $A(t)=A_{0}\sin^{2}[\omega_{0}t/(2N_{c})]\sin(\omega_{0}t)$ for $0\leq{t}\leq2\pi{N_{c}}/\omega_{0}$,
% \begin{equation}
% A(t) = A_{0}\sin^{2}\left(\frac{\omega_{0}t}{2 N_{c}}\right)\sin(\omega_{0}t), \quad (0 \leq t \leq 2\pi N_{c}/\omega_{0}), \label{eq:vecpot}
% \end{equation}
with $\omega_{0}$ the angular frequency (photon energy) and $N_{c}$ the number of cycles.
The laser-driven many-electron system is governed by the time-dependent KS equations
\begin{align}
&i\frac{\partial}{\partial t}\varphi_{j,\sigma}^{}(x,t) \nonumber\\
&= \left\{-\frac{1}{2}\frac{\partial^{2}}{\partial x^{2}} -i A(t)\frac{\partial}{\partial x} + \tilde{v}_{\text{KS}}^{}[\{n_{\sigma}^{}\}](x,t)\right\}\varphi_{j,\sigma}^{}(x,t), \label{eq:dyn_kseq}
\end{align}
where the KS potential $\tilde{v}_{\text{KS}}^{}[\{n_{\sigma}^{}\}](x,t)=v_{\text{ion}}^{}(x)+v_{\text{H}}^{}[n](x,t)+v_{\text{xc}}^{}[\{n_{\sigma}^{}\}](x,t)$
% \begin{equation}
% \tilde{v}_{\text{KS}}^{}[\{n_{\sigma}^{}\}](x,t) = v_{\text{ion}}^{}(x) + v_{\text{H}}^{}[n](x,t) + v_{\text{xc}}^{}[\{n_{\sigma}^{}\}](x,t), \label{eq:dyn_kspot}
% \end{equation}
is time-dependent due to the time dependence of $n(x,t)$ and $n_{\sigma}^{}(x,t)$.
We propagate the time-dependent KS orbitals using the Crank-Nicolson approach with a predictor-corrector step for updating the KS potential \cite{bauer2017book,ullrich2011book}.
The initial conditions for TDDFT calculations, i.e., the field-free ground-state KS orbitals are found via imaginary time propagation with orthogonalization in each time step \cite{bauer2017book}.
The numerical calculations are performed on an equidistant spatial grid with spacing $\Delta x=0.1$ and $21000$ grid points. A fixed step size $\Delta t=0.1$ is used for time propagation, and a convergence check is performed by using $\Delta t=0.05$.

\section{\label{sec:result}Results and Discussion}

In this section, we first describe the influence of doping on the field-free potentials, orbital energies and band structures within the KS scheme.
Then we discuss effects of doping on the HHG spectrum based on the TDDFT calculations, and identify the contribution from a single impurity orbital in the donor-doped case.
Finally, the HHG cutoff is explained by semiclassical analysis.

\subsection{\label{ssec:result1}Doping effects in the DFT description}

\begin{figure}
\includegraphics[width=0.475\textwidth]{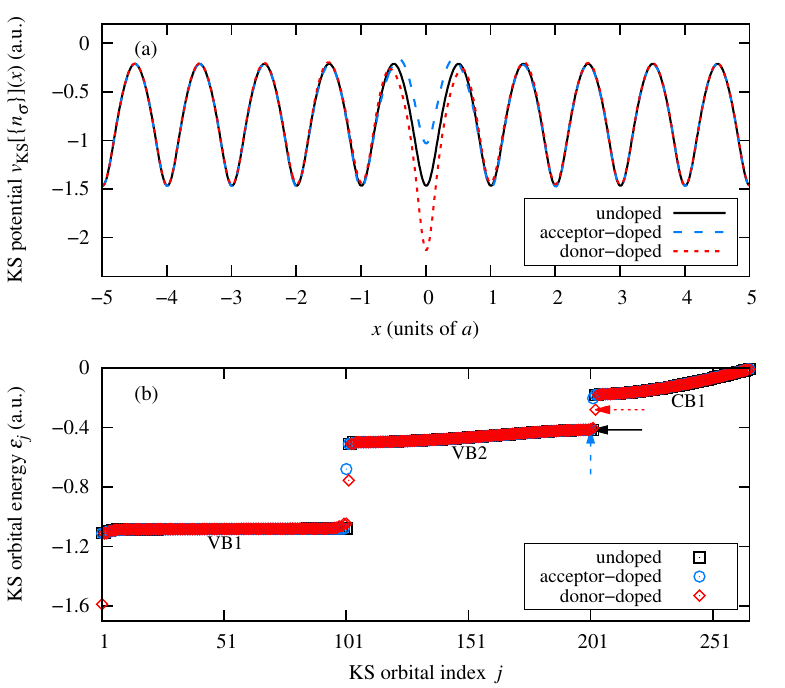}
\caption{Comparisons of undoped and doped systems in the KS scheme. (a) The KS potentials around the impurity for the undoped and doped systems. The separation between two neighboring ions is $a=7$ in our model. (b) The negative-valued KS orbital energies in ascending order, including two valence bands (VB1 and VB2) and part of a conduction band (CB1). For the undoped, acceptor-doped and donor-doped systems, the number of occupied orbitals is $N_{\downarrow}=N_{\uparrow}=202$, $201$ and $203$, respectively (indicated by solid black, dashed blue and dotted red lines with arrows). In the doped cases, isolated orbital energies appear, which cannot be classified into any band.}
\label{fig:fig1}
\end{figure}

We first take a view on the doping-induced change of the field-free properties in the DFT language.
The undoped model was studied in Refs.~\cite{kkh2017pra,kkh2018pra}.
Here we first emphasize the differences between the doped and undoped systems in terms of the static KS potential which is obtained by imaginary time propagation.
The impact of the impurity is restricted in real space to a small region around its position, the center of the system in this case, as shown in Fig.~\ref{fig:fig1}(a).
Compared with the undoped system, the acceptor- and donor-type doping results in a shallower and deeper effective potential around the impurity ion, respectively.
Unlike the single-active-electron approach using a parametrized potential \cite{huang2017pra}, the effective potential in the DFT description is self-consistently found in a many-electron model.
% \footnote{Our self-consistent results also show that the doping-induced change of the effective potential is not exactly restricted to only one unit cell as assumed in Ref.~\cite{huang2017pra}.}.

\begin{figure*}
\includegraphics[width=0.95\textwidth]{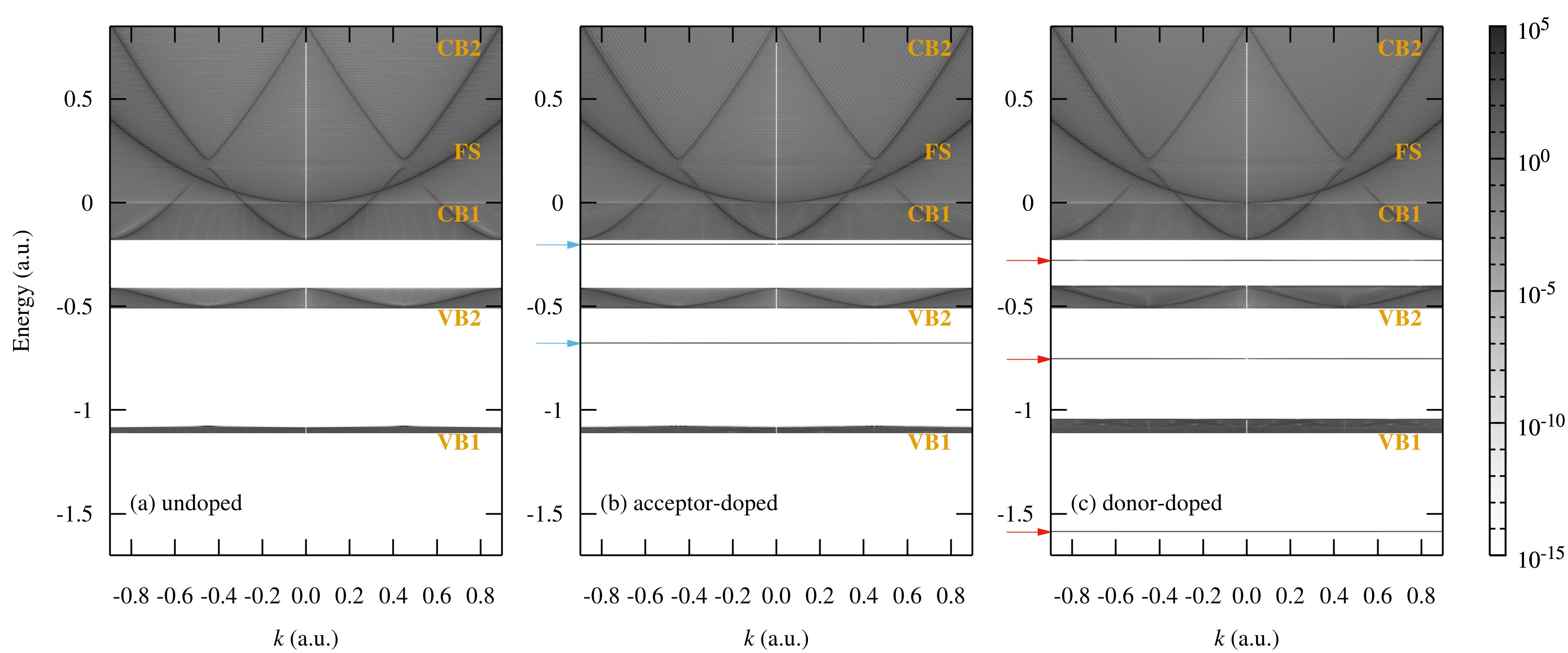}
\caption{Band structures (VB1, VB2, CB1 and CB2) for undoped and doped systems with $N=101$ nuclei, displayed by norm square of the Fourier-transformed KS orbitals. The free-space (FS) dispersion $k^{2}/2$ is also visable due to the finiteness of the simulation box (see text). In the doped cases, the isolated horizontal lines correspond to impurity orbitals, which are also indicated by arrows. The first-Brillouin-zone boundary is at $k=\pi/a=0.449$.}
\label{fig:fig2}
\end{figure*}

With the static KS potential at hand, one can find the occupied and unoccupied orbitals by diagonalization, together with their corresponding energies.
Note that in the KS scheme, the classification into occupied and unoccupied orbitals is automatically determined by the number of electrons with the Pauli exclusion principle satisfied, which is another advantage over the frequently-used single-active-electron approach (see, e.g., Refs.~\cite{hawkins2015pra,wmx2015pra,wmx2016pra,bxb2016pra,bxb2016pra2,bxb2018pra1,bxb2018pra2,huang2017pra,jin2018jpb}).
To illustrate the doping effects on the KS orbitals, we present in Fig.~\ref{fig:fig1}(b) the negative-valued orbital energies in ascending order. % for both the undoped and doped cases.
The energies include two valence bands (VB1 and VB2) and part of a conduction band (CB1), and the ``in-band'' energies remain almost unchanged by doping.
Here we classify the orbital energies into bands, since our model behaves like a solid for a sufficiently large system size \cite{kkh2018pra}.
The visible band gap (BG) allows us to identify the doping-induced impurity orbitals with isolated energies.
In addition to their isolated energies, the impurity orbitals are spatially localized around the impurity ion (see the Appendix).
In the case of acceptor- (donor-) type doping, the impurity orbital energetically located between VB2 and CB1 is unoccupied (occupied).
It is worth noting that the impurity KS orbitals can be seen as a self-consistent DFT description of the ``impurity states'' introduced in the pioneering works \cite{slater1949pr,luttinger1955pr}, see also reviews \cite{impurity1957ssp,impurity1974rpp,impurity1978rmp}.
For our purpose of exploring doping effects on HHG, it is straightforward to perform self-consistent simulations without resorting to additional assumptions.
Thus both ``in-band'' and impurity orbitals are taken into account, and one can investigate their respective contributions in the many-electron processes.
Whether an impurity orbital (state) plays a particular role in HHG will be studied in Sec.~\ref{ssec:result3}.

The band structures can be constructed from the Fourier-transformed orbitals (in $k$-space), as done in Refs.~\cite{kkh2017pra,kkh2018pra}.
Figure~\ref{fig:fig2} shows the norm square of the $k$-space KS orbitals for the undoped and doped systems with $N=101$ nuclei.
The energy range in this plot includes two valence bands (VB1 and VB2) and two conduction band (CB1 and CB2).
Since our simulations are performed in a finite box, the free-space (FS) dispersion $k^{2}/2$ is also present in Fig.~\ref{fig:fig2}.
For HHG in solids, however, the FS parabola does not play any noticeable role \cite{kkh2017pra,kkh2018pra}.
We find that the band structures are almost not affected by doping, except that the energy range of VB1 becomes wider when the system is donor-doped.
The role of VB1 in HHG processes, however, is negligible because of the flat band structure and the low orbital energies, which has been demonstrated in Ref.~\cite{kkh2017pra}.
% In the undoped (or acceptor-doped) system, the orbitals initially in VB2 play a dominant role in the total HHG signal \cite{kkh2017pra,kkh2018pra}.
Figure~\ref{fig:fig2} clearly shows that the considered doping scenarios generate energetically isolated orbitals which do not belong to any band.
One can identify two impurity orbitals from Fig.~\ref{fig:fig2}(b) in the acceptor-type doping case, and three impurity orbitals from Fig.~\ref{fig:fig2}(c) in the donor-type doping case.
A detailed view of these impurity orbitals in real space is presented in the Appendix.

\subsection{\label{ssec:result2}Doping effects on the HHG spectrum}

Using the ground-state occupied KS orbitals as the initial state, we perform TDDFT calculations for the systems interacting with a $6$-cycle laser pulse of frequency $\omega_{0}=0.0114$ corresponding to a wavelength of $\sim4000$ nm.
We compute the time-dependent current
\begin{equation}
J(t) = \sum_{j,\sigma}^{} \int dx \ \text{Re}\left[\varphi_{j,\sigma}^{*}(x,t)\left(-i\frac{\partial}{\partial x}+A(t)\right)\varphi_{j,\sigma}^{}(x,t) \right], \label{eq:current}
\end{equation}
% \begin{equation}
% J(t) = \sum_{j,\sigma}^{} \int dx \ \text{Im}\left[\varphi_{j,\sigma}^{*}(x,t)\frac{\partial}{\partial x}\varphi_{j,\sigma}^{}(x,t) \right], \label{eq:current_na}
% \end{equation}
and evaluate the HHG spectral intensity as the modulus square of the Fourier-transformed current, i.e., $S(\omega)\propto|\int{dt}J(t)\exp(-i{\omega}t)|^{2}$.
Here we do not account for macroscopic propagation effects, which may modify the HHG spectra via absorption and phase mismatch.
Such propagation effects, however, can be mitigated by controlling the thickness of target materials \cite{ndabashimiye2016nature}. % or using nanowires \cite{cmcdo2017prl2}.
Therefore we expect our conclusions to hold for a thin target material.

\begin{figure}
\includegraphics[width=0.475\textwidth]{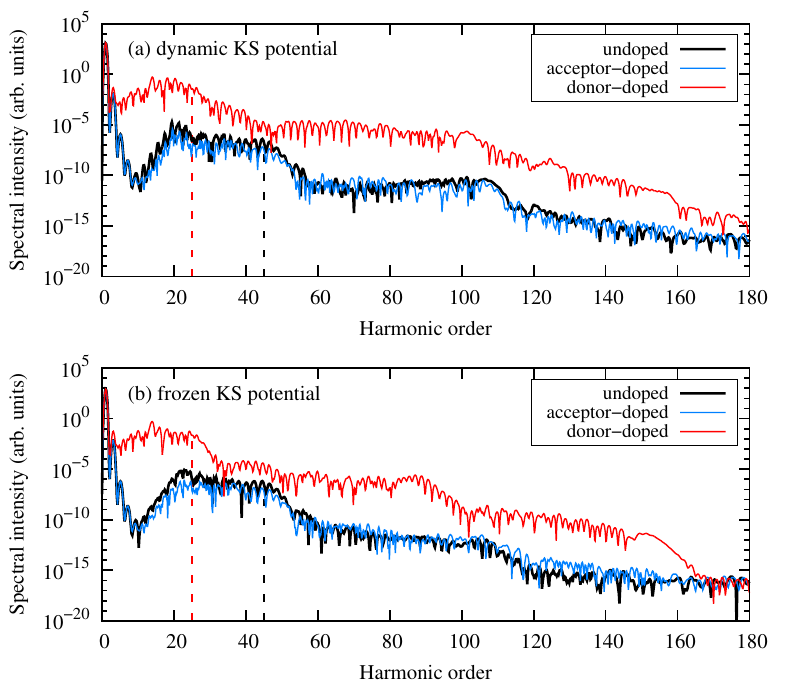}
\caption{HHG spectra for undoped and doped systems obtained from (a) TDDFT calculations with the KS potential dynamically updated according to the time-dependent densities, and (b) calculations with the KS potential frozen to its field-free ground-state form. In both panels, the donor-doped system gives the highest yield.
The laser parameters correspond to a wavelength of $\sim4000$ nm and a peak intensity of $\sim2.2\times10^{11}$ W/cm$^{2}$ (see text). The vertical dashed lines indicate the first cutoffs: a cutoff of order 45 is observed for the undoped and acceptor-doped systems while a cutoff of order 25 is observed for the donor-doped system.} % (see semiclassical analysis below).}
\label{fig:fig3}
\end{figure}

Figure~\ref{fig:fig3}(a) shows HHG spectra for the undoped and doped systems obtained from TDDFT calculations, for $A_{0}=0.22$ which corresponds to an intensity of $\sim2.2\times10^{11}$ W/cm$^{2}$. The BG between CB1 and VB2 is $0.235\sim6.4$ eV typical for a dielectric, implying that harmonics up to order 20 are in the sub-BG regime for the undoped system.
The spectral minimum in the sub-BG region stems from the fact that the intensity of intraband harmonics decreases with increasing order %\cite{intra2018prl}
and the interband harmonics become dominant when going into the above-BG regime.
The HHG spectrum for the acceptor-doped system is very similar to that for the undoped system.
In contrast, the overall HHG for the donor-doped system is enhanced by $\sim2-6$ orders of magnitude.
This significant enhancement of the HHG efficiency would be favorable for applications as a coherent source of VUV and XUV radiation.
We also perform calculations with a frozen ground-state KS potential.
Such a frozen-KS-potential approach is applicable for moderate intensities well below the damage threshold of solids \cite{kkh2017pra}.
The spectra obtained from these calculations are shown in Fig.~\ref{fig:fig3}(b).
The considerable enhancement of HHG in the donor-type doping case and the similarity of the undoped and acceptor-doped cases are also found with this approach.
As will be shown in Sec.~\ref{ssec:result3}, the frozen-KS-potential approach offers the possibility to identify the contribution from a single impurity orbital, which provides insights into the observed enhancement of HHG in the donor-type doping case.
In addition, a different cutoff for the donor-doped system is found in Fig.~\ref{fig:fig3}, which will be analyzed in Sec.~\ref{ssec:result4}.

\subsection{\label{ssec:result3}Role of a single impurity orbital}

To understand the doping effects on HHG, we link our findings in Fig.~\ref{fig:fig3} to the doping-induced changes of the field-free properties displayed in Fig.~\ref{fig:fig1} and Fig.~\ref{fig:fig2}.
Let us first revisit the intra- and interband contributions of HHG in undoped solids \cite{vampa2014prl,vampa2017jpb}.
Intraband HHG stems from the laser-driven electron motion in bands due to the anharmonicity of the band structure.
Interband HHG is described by the generalized three-step model for band-gap materials: first an electron tunnels into the conduction band, leaving a hole in the valence band; then the electron and hole move in their respective bands and may recombine at a later time, emitting a photon with energy above the BG.
Thus the BG energy plays a similar role as the ionization potential in atomic HHG.
If the energy of the highest occupied orbital is close to the lowest conduction-band energy, the electron has a high probability to tunnel into the conduction band, since the tunneling rate is exponentially sensitive to the energy gap \cite{keldysh1965jetp}.
As shown in Fig.~\ref{fig:fig1}(b) and Fig.~\ref{fig:fig2}, the considered doping causes no obvious change to the band structures, except for introducing the impurity orbitals.
The similarity of HHG spectra for the acceptor-doped and undoped systems can then be understood by noting that the highest occupied orbital in both cases is at the top of VB2 [see Fig.~\ref{fig:fig1}(b)].
For the donor-doped system, we expect that the highest occupied impurity orbital within the BG is responsible for the enhancement of HHG observed in Fig.~\ref{fig:fig3}.

\begin{figure}
\includegraphics[width=0.475\textwidth]{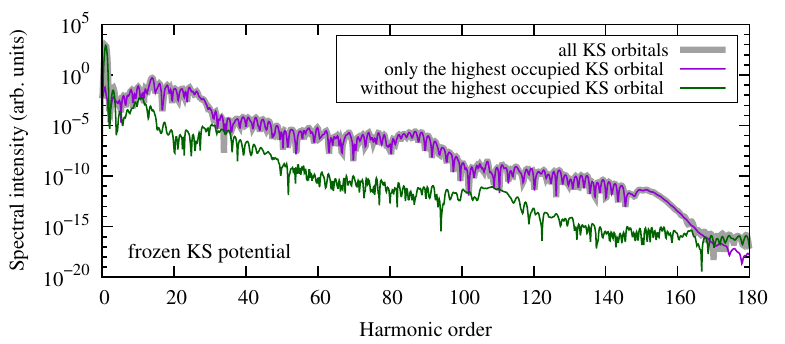}
\caption{HHG spectra for the donor-doped system obtained from calculations with the frozen KS potential [the thick grey curve, which is the same as the upper red curve in Fig.~\ref{fig:fig3}(b)]. The thin purple curve and the thin green curve are obtained by Fourier transforming the partial current calculated with only the highest occupied KS orbital and without this orbital, respectively.}
\label{fig:fig4}
\end{figure}

The role of the highest occupied impurity orbital in the donor-doped system can be highlighted within the frozen-KS-potential approach.
To this end, we calculate the current from the highest occupied orbital by restricting the sum in Eq.~\eqref{eq:current} to that orbital, and compare the resulting HHG spectrum with the total one in Fig.~\ref{fig:fig4}.
One can see that for a wide range of harmonic orders, from $\sim10$ to $\sim160$, the contribution from the single impurity orbital agrees with the total spectrum.
Therefore we attribute the enhancement of HHG in the donor-doped system to the highest occupied impurity orbital that has an isolated energy within the BG.
% Note that the important role of the highest occupied impurity orbital can only be inferred with the frozen-KS-potential approach.
% In full TDDFT calculations, the time-dependent change of the KS potential affects all the orbitals, mixing their contributions in the total current.
We note that impurity-state HHG was modeled in a recent work \cite{impurity2018prb} taking only the impurity-state contribution into account when evaluating the HHG spectrum.
Based on the self-consistent many-electron calculations, our present work evidences that the impurity-state HHG signal may agree with the total signal for a wide range of harmonic orders in a donor-doped band-gap material.

\subsection{\label{ssec:result4}Semiclassical analysis of the cutoff}

\begin{figure}
\includegraphics[width=0.475\textwidth]{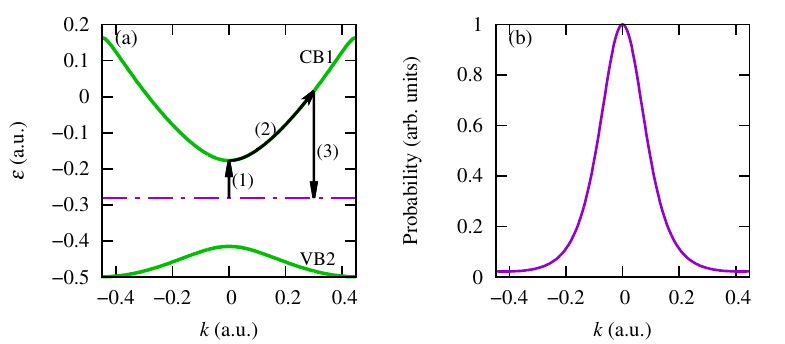}
\caption{(a) A three-step model for HHG from the impurity-state electron, illustrated in $k$-space. The horizontal dot-dashed line indicates the donor state within the BG, i.e., the highest occupied impurity orbital for the donor-doped system. The (green) curves show the band structures of CB1 and VB2. The three steps for HHG are indicated by (black) lines with arrows. For the considered system and laser frequency, it requires at least 9 photons to excite the impurity-state electron to CB1, which is in the tunneling regime. (b) The $k$-space probability distribution of the impurity orbital depicted in (a). The first-Brillouin-zone boundary is at $k=\pi/a=0.449$.}
\label{fig:fig5}
\end{figure}

Having demonstrated that HHG from the highest occupied impurity orbital is dominant for the donor-doped system, we now elucidate the corresponding mechanism by a semiclassical analysis.
HHG from the donor-state electron can be described by a three-step model: the impurity-state electron tunnels into the conduction band, moving according to the band structure, and recombines with the impurity state when driven back by the external field.
This mechanism is illustrated in $k$-space in Fig.~\ref{fig:fig5}(a) \footnote{For the considered vector-potential amplitude $A_{0}=0.22<\pi/(2a)$, we simply restrict to a single conduction band, CB1, since the semiclassical electron will not reach the first-Brillouin-zone boundary [Eq.~\eqref{eq:cbmotion}].}.
Here the band-structure curves are extracted from Fig.~\ref{fig:fig2} and fitted to be continuous, as done in Ref.~\cite{kkh2018pra}.
The impurity state is depicted as a completely flat band, since its $k$-space distribution spreads over the first Brillouin zone [see Fig.~\ref{fig:fig5}(b)].
Similarly to the semiclassical model of interband HHG in solids \cite{vampa2015prb,vampa2017jpb}, we estimate the cutoff for HHG from the impurity-state electron.
First, the tunneling step is considered to occur at $k_{0}=0$ corresponding to the minimum of the conduction band energy.
With the tunneling time denoted by $t_{i}$, the electron trajectory after tunneling is given as \cite{vampa2015prb,vampa2017jpb}
\begin{equation}
x(t)-x_{i} = \int_{t_{i}}^{t}d\tau \nabla_{k}\mathcal{E}_{c}[k(\tau)],\quad k(\tau) = A(\tau)-A(t_{i}), \label{eq:cbmotion}
\end{equation}
where $\mathcal{E}_{c}[k(\tau)]$ is the band structure of CB1 and $x_{i}=0$ is the initial position of the electron.
When the electron returns to its initial position at time $t_{r}$, it recombines to the impurity state, emitting an energy of $\mathcal{E}_{c}[k(t_{r})]-\varepsilon_{i}$ with $\varepsilon_{i}$ the impurity-state energy.
Since there is no hole motion in the valence band, the three-step model for impurity-state HHG is more atomiclike than that for solid HHG.
The only difference compared with the three-step model for atomic HHG is that the electron motion after tunneling is determined by the band structure $\mathcal{E}_{c}(k)$, rather than the free-space dispersion relation $\mathcal{E}(k)=k^{2}/2$.

\begin{figure}
\includegraphics[width=0.475\textwidth]{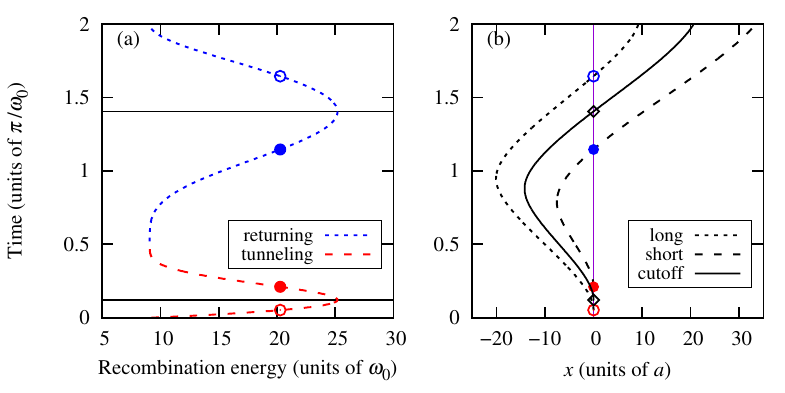}
\caption{(a) The mapping between recombination energies (in units of $\omega_{0}=0.0114$) and semiclassical trajectories (characterized by tunneling and returning times), within one cycle of a sinusoidal vector potential with amplitude $A_{0}=0.22$. Zero on the vertical axis means a time at which the vector potential is 0. A recombination energy is associated with a pair of trajectories (referred to as long and short, in terms of the time difference between tunneling and returning). An example of long (short) trajectories is indicated by the empty (filled) circles. The horizontal lines indicate a pair of tunneling and returning times that produce the maximum recombination energy, i.e., the cutoff trajectory. (b) A real-space view of the cutoff trajectory and the example pair of long and short trajectories indicated in (a). The tunneling and returning times are marked on the vertical line $x=0$ corresponding to the impurity center. The separation between two neighboring nuclei is $a=7$ in our model.}
\label{fig:fig6}
\end{figure}

By searching for the maximum recombination energy with different tunneling times taken into consideration, we estimate the cutoff of impurity-state HHG.
Note that the trajectories in the semiclassical analysis are characterized by tunneling and returning times, and a recombination energy is associated with a pair of trajectories (referred to as long and short, in terms of the time difference between tunneling and returning of the electron).
As an illustration, we show in Fig.~\ref{fig:fig6}(a) the mapping between recombination energies and trajectories, for the considered laser parameters.
The maximum recombination energy found by the semiclassical analysis corresponds to the $25$-th harmonic, which agrees well with the observed cutoff for the donor-doped system in Fig.~\ref{fig:fig3}.
We also present in Fig.~\ref{fig:fig6}(b) a real-space view of the cutoff trajectory and a pair of long and short trajectories.
One can see that the impurity-state electron can move many unit-cell distances away from the impurity ion.

We mention in passing that a cutoff analysis based on the three-step model for solid HHG was also performed for the undoped system, and the corresponding cutoff estimated from the semiclassical electron-hole dynamics also agrees with that observed for the undoped system in Fig.~\ref{fig:fig3}.
Here we focus on the first cutoff in the semiclassical analysis, since the harmonics up to the first cutoff are of more practical interest due to their relatively stronger signals.
Explanation for higher cutoffs would require a model accounting for more complicated processes, which is beyond the scope of this work.
So far the harmonics beyond the first cutoff are seldom measured experimentally \cite{ndabashimiye2016nature}. Our work indicates that experimental observation of such high-order harmonics might be less difficult for donor-doped materials.

\section{\label{sec:concl}Conclusion}

In summary, we found that a donor-doped band-gap material can produce HHG much more efficiently than undoped and acceptor-doped materials.
This significant enhancement of HHG stems from the highest occupied impurity state with an isolated energy within the BG. %, which has a higher probability to tunnel into the conduction band than the valence band states.
In contrast to HHG described by electron-hole dynamics in undoped solids, HHG from the impurity-state electron is more atomiclike, i.e., the electron moving in the conduction band will recombine with the impurity state rather than a moving hole in the valence band.
The mechanism of the impurity-state HHG can be described by a three-step model where the impurity-state electron tunnels into the conduction band and then moves according to the conduction band structure until recombination.
This leads to a harmonic cutoff different from that in the undoped case, which can be explained by semiclassical analysis based on the band structure.
Our present work implies that donor-doped band-gap materials would be suitable for efficient generation of coherent VUV and XUV radiation.
Exploring ultrafast processes in doped materials with HHG will be interesting for future work. %, which may provide further understanding in the novel research area where strong-field physics meets condensed matter.

\begin{acknowledgments}
We thank Dieter Bauer for making the \mbox{Qprop} code available, on which the TDDFT calculations are based.
L.B.M. acknowledges discussions with Peter Balling and Brian Julsgaard.
This work was supported by the Villum Kann Rasmussen (VKR) Center of Excellence \mbox{QUSCOPE} \--- Quantum Scale Optical Processes. The numerical results were obtained at the Centre for Scientific Computing, Aarhus.
\end{acknowledgments}

\appendix*
\section{\label{sec:append}Impurity orbitals in real space}

\begin{figure*}
\includegraphics[width=0.95\textwidth]{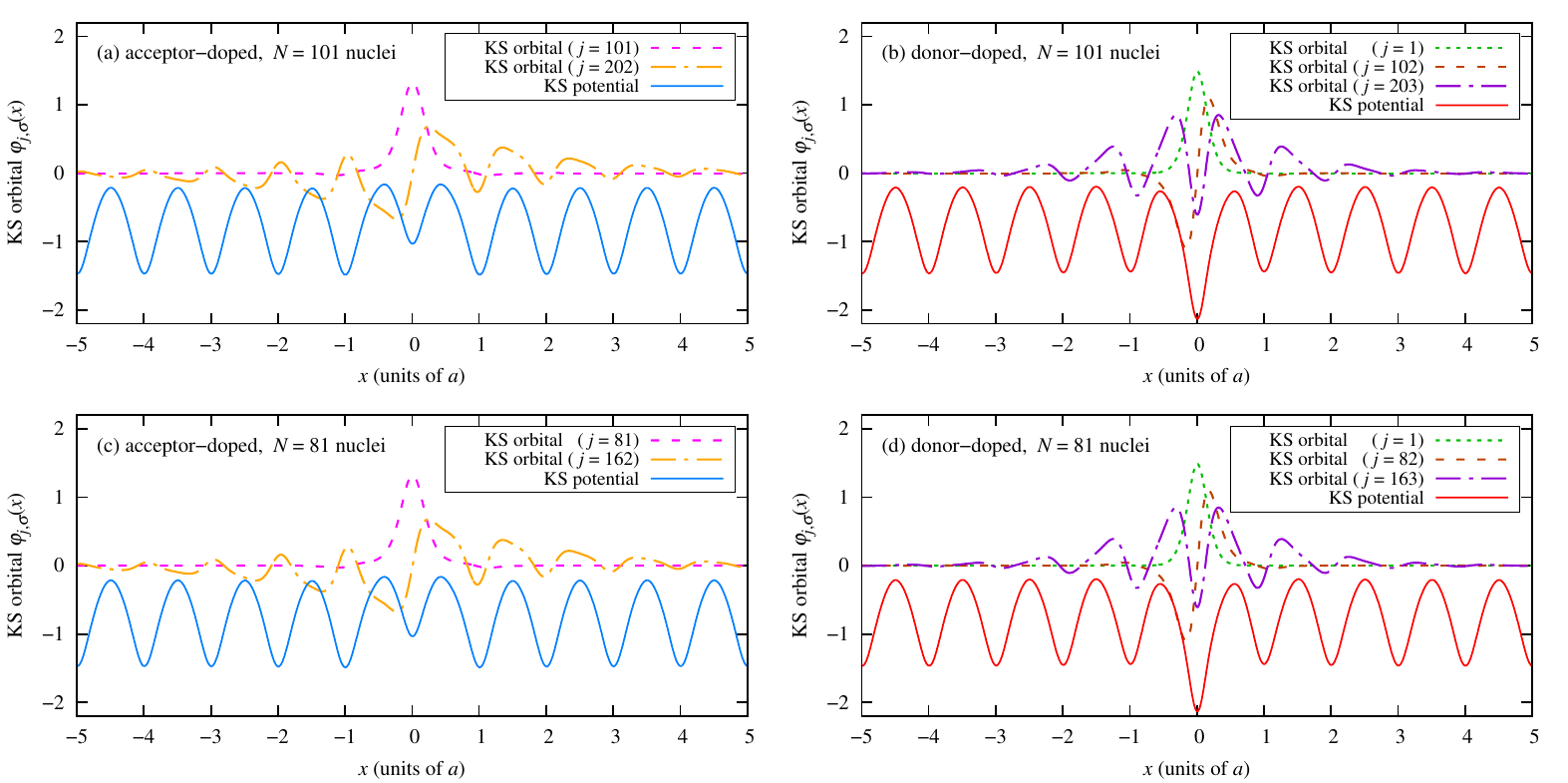}
\caption{(a) A real-space view of the energy-isolated orbitals for the acceptor-doped system with $N=101$ nuclei. The orbitals with index $j=101$ and $j=202$ correspond to the lower and upper isolated energies in Fig.~\ref{fig:fig2}(b), respectively. (b) A real-space view of the energy-isolated orbitals for the donor-doped system. The orbitals with index $j=1$, $j=102$ and $j=203$ correspond to the isolated energies marked by arrows in Fig.~\ref{fig:fig2}(c), from bottom to top. In both acceptor- and donor-type doping cases, the corresponding KS potentials are shown for an illustrative view of the impurity and its neighboring ions. The lower panels (c) and (d) present the impurity orbitals for smaller doped systems with $N=81$ nuclei. In our model, the separation between two neighboring nuclei, i.e., the unit-cell length, is $a=7$.}
\label{fig:fig7}
\end{figure*}

The KS orbitals obtained from the field-free calculations can be chosen to be real-valued.
In Figs.~\ref{fig:fig7}(a) and (b), we provide a real-space view of the energy-isolated orbitals identified from Figs.~\ref{fig:fig2}(b) and (c). 
For the considered model with $N=101$ nuclei, the number of occupied orbitals is $N_{\downarrow}=N_{\uparrow}=202$, $201$ and $203$ for the undoped, acceptor-doped and donor-doped systems, respectively.
Thus the orbital with index $j=202$ in Fig.~\ref{fig:fig7}(a) is the lowest-unoccupied for the acceptor-doped system while the orbital with index $j=203$ in Fig.~\ref{fig:fig7}(b) is the highest occupied for the donor-doped system.
One can see that the impurity orbitals are located around the impurity ion, spreading over a few neighboring ions.

We also show the impurity orbitals for smaller doped systems with $N=81$ nuclei in Figs.~\ref{fig:fig7}(c) and (d).
Compared with Figs.~\ref{fig:fig7}(a) and (b), this variation of the system size does not cause any discernible change of the KS potential and the impurity orbitals.
Note that for the model with $N=81$ nuclei, the number of occupied orbitals is $N_{\downarrow}=N_{\uparrow}=162$, $161$ and $163$ for the undoped, acceptor-doped and donor-doped systems, respectively.
The indexes of the impurity orbitals in Figs.~\ref{fig:fig7}(c) and (d) are therefore different from those in Figs.~\ref{fig:fig7}(a) and (b).
The energies of the impurity orbitals are, however, not affected, because of the negligible change of the KS potential and the impurity orbitals.
In the frozen-KS-potential approach, the HHG processes are simulated by propagating the KS orbitals in the frozen KS potential, and it is shown in Fig.~\ref{fig:fig4} that the highest occupied impurity-state orbital determines the total HHG spectrum for a wide range of harmonic orders in the case of donor-type doping.
Thus we infer that our discussion of the impurity effects is insensitive to the system size considered in our simulations, as long as the system is sufficiently large (e.g., with more than $\sim80$ nuclei) such that it behaves like a solid and contains the real-space motion of the impurity-state electron.

Although our discussion of doping effects is based on TDDFT simulations in a model, the underlying physics of the impurity-state HHG should be true for a real band-gap material.
Since the doping-induced impurity orbitals are spatially localized around the impurity ion, we expect our results to be qualitatively valid for a target system with more than one impurity ion, as long as the impurity ions are many unit-cell distances away from each other.

\clearpage

% \bibliographystyle{apsrev4-1}
% \bibliography{refs_rev}

\begin{thebibliography}{51}%
\makeatletter
\providecommand \@ifxundefined [1]{%
 \@ifx{#1\undefined}
}%
\providecommand \@ifnum [1]{%
 \ifnum #1\expandafter \@firstoftwo
 \else \expandafter \@secondoftwo
 \fi
}%
\providecommand \@ifx [1]{%
 \ifx #1\expandafter \@firstoftwo
 \else \expandafter \@secondoftwo
 \fi
}%
\providecommand \natexlab [1]{#1}%
\providecommand \enquote  [1]{``#1''}%
\providecommand \bibnamefont  [1]{#1}%
\providecommand \bibfnamefont [1]{#1}%
\providecommand \citenamefont [1]{#1}%
\providecommand \href@noop [0]{\@secondoftwo}%
\providecommand \href [0]{\begingroup \@sanitize@url \@href}%
\providecommand \@href[1]{\@@startlink{#1}\@@href}%
\providecommand \@@href[1]{\endgroup#1\@@endlink}%
\providecommand \@sanitize@url [0]{\catcode `\\12\catcode `\$12\catcode
  `\&12\catcode `\#12\catcode `\^12\catcode `\_12\catcode `\%12\relax}%
\providecommand \@@startlink[1]{}%
\providecommand \@@endlink[0]{}%
\providecommand \url  [0]{\begingroup\@sanitize@url \@url }%
\providecommand \@url [1]{\endgroup\@href {#1}{\urlprefix }}%
\providecommand \urlprefix  [0]{URL }%
\providecommand \Eprint [0]{\href }%
\providecommand \doibase [0]{http://dx.doi.org/}%
\providecommand \selectlanguage [0]{\@gobble}%
\providecommand \bibinfo  [0]{\@secondoftwo}%
\providecommand \bibfield  [0]{\@secondoftwo}%
\providecommand \translation [1]{[#1]}%
\providecommand \BibitemOpen [0]{}%
\providecommand \bibitemStop [0]{}%
\providecommand \bibitemNoStop [0]{.\EOS\space}%
\providecommand \EOS [0]{\spacefactor3000\relax}%
\providecommand \BibitemShut  [1]{\csname bibitem#1\endcsname}%
\let\auto@bib@innerbib\@empty
%</preamble>
\bibitem [{\citenamefont {Krause}\ \emph {et~al.}(1992)\citenamefont {Krause},
  \citenamefont {Schafer},\ and\ \citenamefont {Kulander}}]{hhg1992prl}%
  \BibitemOpen
  \bibfield  {author} {\bibinfo {author} {\bibfnamefont {J.~L.}\ \bibnamefont
  {Krause}}, \bibinfo {author} {\bibfnamefont {K.~J.}\ \bibnamefont {Schafer}},
  \ and\ \bibinfo {author} {\bibfnamefont {K.~C.}\ \bibnamefont {Kulander}},\
  }\enquote {\bibinfo {title} {High-order harmonic generation from atoms and
  ions in the high intensity regime},}\ \href {\doibase
  10.1103/PhysRevLett.68.3535} {\bibfield  {journal} {\bibinfo  {journal}
  {Phys. Rev. Lett.}\ }\textbf {\bibinfo {volume} {68}},\ \bibinfo {pages}
  {3535} (\bibinfo {year} {1992})}\BibitemShut {NoStop}%
\bibitem [{\citenamefont {L'Huillier}\ and\ \citenamefont
  {Balcou}(1993)}]{hhg1993prl}%
  \BibitemOpen
  \bibfield  {author} {\bibinfo {author} {\bibfnamefont {A.}~\bibnamefont
  {L'Huillier}}\ and\ \bibinfo {author} {\bibfnamefont {P.}~\bibnamefont
  {Balcou}},\ }\enquote {\bibinfo {title} {High-order harmonic generation in
  rare gases with a 1-ps 1053-nm laser},}\ \href {\doibase
  10.1103/PhysRevLett.70.774} {\bibfield  {journal} {\bibinfo  {journal} {Phys.
  Rev. Lett.}\ }\textbf {\bibinfo {volume} {70}},\ \bibinfo {pages} {774}
  (\bibinfo {year} {1993})}\BibitemShut {NoStop}%
\bibitem [{\citenamefont {Krausz}\ and\ \citenamefont
  {Ivanov}(2009)}]{atto2009rmp}%
  \BibitemOpen
  \bibfield  {author} {\bibinfo {author} {\bibfnamefont {F.}~\bibnamefont
  {Krausz}}\ and\ \bibinfo {author} {\bibfnamefont {M.}~\bibnamefont
  {Ivanov}},\ }\enquote {\bibinfo {title} {Attosecond physics},}\ \href
  {\doibase 10.1103/RevModPhys.81.163} {\bibfield  {journal} {\bibinfo
  {journal} {Rev. Mod. Phys.}\ }\textbf {\bibinfo {volume} {81}},\ \bibinfo
  {pages} {163} (\bibinfo {year} {2009})}\BibitemShut {NoStop}%
\bibitem [{\citenamefont {Ghimire}\ \emph {et~al.}(2011)\citenamefont
  {Ghimire}, \citenamefont {DiChiara}, \citenamefont {Sistrunk}, \citenamefont
  {Agostini}, \citenamefont {DiMauro},\ and\ \citenamefont
  {Reis}}]{ghimire2011natphys}%
  \BibitemOpen
  \bibfield  {author} {\bibinfo {author} {\bibfnamefont {S.}~\bibnamefont
  {Ghimire}}, \bibinfo {author} {\bibfnamefont {A.~D.}\ \bibnamefont
  {DiChiara}}, \bibinfo {author} {\bibfnamefont {E.}~\bibnamefont {Sistrunk}},
  \bibinfo {author} {\bibfnamefont {P.}~\bibnamefont {Agostini}}, \bibinfo
  {author} {\bibfnamefont {L.~F.}\ \bibnamefont {DiMauro}}, \ and\ \bibinfo
  {author} {\bibfnamefont {D.~A.}\ \bibnamefont {Reis}},\ }\enquote {\bibinfo
  {title} {Observation of high-order harmonic generation in a bulk crystal},}\
  \href {\doibase 10.1038/nphys1847} {\bibfield  {journal} {\bibinfo  {journal}
  {Nat. Phys.}\ }\textbf {\bibinfo {volume} {7}},\ \bibinfo {pages} {138}
  (\bibinfo {year} {2011})}\BibitemShut {NoStop}%
\bibitem [{\citenamefont {Schubert}\ \emph {et~al.}(2014)\citenamefont
  {Schubert}, \citenamefont {Hohenleutner}, \citenamefont {Langer},
  \citenamefont {Urbanek}, \citenamefont {Lange}, \citenamefont {Huttner},
  \citenamefont {Golde}, \citenamefont {Meier}, \citenamefont {Kira},
  \citenamefont {Koch} \emph {et~al.}}]{schubert2014natphoton}%
  \BibitemOpen
  \bibfield  {author} {\bibinfo {author} {\bibfnamefont {O.}~\bibnamefont
  {Schubert}}, \bibinfo {author} {\bibfnamefont {M.}~\bibnamefont
  {Hohenleutner}}, \bibinfo {author} {\bibfnamefont {F.}~\bibnamefont
  {Langer}}, \bibinfo {author} {\bibfnamefont {B.}~\bibnamefont {Urbanek}},
  \bibinfo {author} {\bibfnamefont {C.}~\bibnamefont {Lange}}, \bibinfo
  {author} {\bibfnamefont {U.}~\bibnamefont {Huttner}}, \bibinfo {author}
  {\bibfnamefont {D.}~\bibnamefont {Golde}}, \bibinfo {author} {\bibfnamefont
  {T.}~\bibnamefont {Meier}}, \bibinfo {author} {\bibfnamefont
  {M.}~\bibnamefont {Kira}}, \bibinfo {author} {\bibfnamefont {S.~W.}\
  \bibnamefont {Koch}},  \emph {et~al.},\ }\enquote {\bibinfo {title}
  {Sub-cycle control of terahertz high-harmonic generation by dynamical Bloch
  oscillations},}\ \href {\doibase 10.1038/nphoton.2013.349} {\bibfield
  {journal} {\bibinfo  {journal} {Nat. Photon.}\ }\textbf {\bibinfo {volume}
  {8}},\ \bibinfo {pages} {119} (\bibinfo {year} {2014})}\BibitemShut {NoStop}%
\bibitem [{\citenamefont {Luu}\ \emph {et~al.}(2015)\citenamefont {Luu},
  \citenamefont {Garg}, \citenamefont {Kruchinin}, \citenamefont {Moulet},
  \citenamefont {Hassan},\ and\ \citenamefont {Goulielmakis}}]{luu2015nature}%
  \BibitemOpen
  \bibfield  {author} {\bibinfo {author} {\bibfnamefont {T.~T.}\ \bibnamefont
  {Luu}}, \bibinfo {author} {\bibfnamefont {M.}~\bibnamefont {Garg}}, \bibinfo
  {author} {\bibfnamefont {S.~Y.}\ \bibnamefont {Kruchinin}}, \bibinfo {author}
  {\bibfnamefont {A.}~\bibnamefont {Moulet}}, \bibinfo {author} {\bibfnamefont
  {M.~T.}\ \bibnamefont {Hassan}}, \ and\ \bibinfo {author} {\bibfnamefont
  {E.}~\bibnamefont {Goulielmakis}},\ }\enquote {\bibinfo {title} {Extreme
  ultraviolet high-harmonic spectroscopy of solids},}\ \href {\doibase
  10.1038/nature14456} {\bibfield  {journal} {\bibinfo  {journal} {Nature}\
  }\textbf {\bibinfo {volume} {521}},\ \bibinfo {pages} {498} (\bibinfo {year}
  {2015})}\BibitemShut {NoStop}%
\bibitem [{\citenamefont {Vampa}\ \emph
  {et~al.}(2015{\natexlab{a}})\citenamefont {Vampa}, \citenamefont {Hammond},
  \citenamefont {Thir{\'e}}, \citenamefont {Schmidt}, \citenamefont
  {L{\'e}gar{\'e}}, \citenamefont {McDonald}, \citenamefont {Brabec},\ and\
  \citenamefont {Corkum}}]{vampa2015nature}%
  \BibitemOpen
  \bibfield  {author} {\bibinfo {author} {\bibfnamefont {G.}~\bibnamefont
  {Vampa}}, \bibinfo {author} {\bibfnamefont {T.}~\bibnamefont {Hammond}},
  \bibinfo {author} {\bibfnamefont {N.}~\bibnamefont {Thir{\'e}}}, \bibinfo
  {author} {\bibfnamefont {B.}~\bibnamefont {Schmidt}}, \bibinfo {author}
  {\bibfnamefont {F.}~\bibnamefont {L{\'e}gar{\'e}}}, \bibinfo {author}
  {\bibfnamefont {C.}~\bibnamefont {McDonald}}, \bibinfo {author}
  {\bibfnamefont {T.}~\bibnamefont {Brabec}}, \ and\ \bibinfo {author}
  {\bibfnamefont {P.}~\bibnamefont {Corkum}},\ }\enquote {\bibinfo {title}
  {Linking high harmonics from gases and solids},}\ \href {\doibase
  10.1038/nature14517} {\bibfield  {journal} {\bibinfo  {journal} {Nature}\
  }\textbf {\bibinfo {volume} {522}},\ \bibinfo {pages} {462} (\bibinfo {year}
  {2015}{\natexlab{a}})}\BibitemShut {NoStop}%
\bibitem [{\citenamefont {Hohenleutner}\ \emph {et~al.}(2015)\citenamefont
  {Hohenleutner}, \citenamefont {Langer}, \citenamefont {Schubert},
  \citenamefont {Knorr}, \citenamefont {Huttner}, \citenamefont {Koch},
  \citenamefont {Kira},\ and\ \citenamefont {Huber}}]{hohenleutner2015nature}%
  \BibitemOpen
  \bibfield  {author} {\bibinfo {author} {\bibfnamefont {M.}~\bibnamefont
  {Hohenleutner}}, \bibinfo {author} {\bibfnamefont {F.}~\bibnamefont
  {Langer}}, \bibinfo {author} {\bibfnamefont {O.}~\bibnamefont {Schubert}},
  \bibinfo {author} {\bibfnamefont {M.}~\bibnamefont {Knorr}}, \bibinfo
  {author} {\bibfnamefont {U.}~\bibnamefont {Huttner}}, \bibinfo {author}
  {\bibfnamefont {S.}~\bibnamefont {Koch}}, \bibinfo {author} {\bibfnamefont
  {M.}~\bibnamefont {Kira}}, \ and\ \bibinfo {author} {\bibfnamefont
  {R.}~\bibnamefont {Huber}},\ }\enquote {\bibinfo {title} {Real-time
  observation of interfering crystal electrons in high-harmonic generation},}\
  \href {\doibase 10.1038/nature14652} {\bibfield  {journal} {\bibinfo
  {journal} {Nature}\ }\textbf {\bibinfo {volume} {523}},\ \bibinfo {pages}
  {572} (\bibinfo {year} {2015})}\BibitemShut {NoStop}%
\bibitem [{\citenamefont {Ndabashimiye}\ \emph {et~al.}(2016)\citenamefont
  {Ndabashimiye}, \citenamefont {Ghimire}, \citenamefont {Wu}, \citenamefont
  {Browne}, \citenamefont {Schafer}, \citenamefont {Gaarde},\ and\
  \citenamefont {Reis}}]{ndabashimiye2016nature}%
  \BibitemOpen
  \bibfield  {author} {\bibinfo {author} {\bibfnamefont {G.}~\bibnamefont
  {Ndabashimiye}}, \bibinfo {author} {\bibfnamefont {S.}~\bibnamefont
  {Ghimire}}, \bibinfo {author} {\bibfnamefont {M.}~\bibnamefont {Wu}},
  \bibinfo {author} {\bibfnamefont {D.~A.}\ \bibnamefont {Browne}}, \bibinfo
  {author} {\bibfnamefont {K.~J.}\ \bibnamefont {Schafer}}, \bibinfo {author}
  {\bibfnamefont {M.~B.}\ \bibnamefont {Gaarde}}, \ and\ \bibinfo {author}
  {\bibfnamefont {D.~A.}\ \bibnamefont {Reis}},\ }\enquote {\bibinfo {title}
  {Solid-state harmonics beyond the atomic limit},}\ \href {\doibase
  10.1038/nature17660} {\bibfield  {journal} {\bibinfo  {journal} {Nature}\
  }\textbf {\bibinfo {volume} {534}},\ \bibinfo {pages} {520} (\bibinfo {year}
  {2016})}\BibitemShut {NoStop}%
\bibitem [{\citenamefont {Kruchinin}\ \emph {et~al.}(2018)\citenamefont
  {Kruchinin}, \citenamefont {Krausz},\ and\ \citenamefont
  {Yakovlev}}]{attosolid2018rmp}%
  \BibitemOpen
  \bibfield  {author} {\bibinfo {author} {\bibfnamefont {S.~Y.}\ \bibnamefont
  {Kruchinin}}, \bibinfo {author} {\bibfnamefont {F.}~\bibnamefont {Krausz}}, \
  and\ \bibinfo {author} {\bibfnamefont {V.~S.}\ \bibnamefont {Yakovlev}},\
  }\enquote {\bibinfo {title} {Colloquium: Strong-field phenomena in periodic
  systems},}\ \href {\doibase 10.1103/RevModPhys.90.021002} {\bibfield
  {journal} {\bibinfo  {journal} {Rev. Mod. Phys.}\ }\textbf {\bibinfo {volume}
  {90}},\ \bibinfo {pages} {021002} (\bibinfo {year} {2018})}\BibitemShut
  {NoStop}%
\bibitem [{\citenamefont {Corkum}(1993)}]{threestep1993prl}%
  \BibitemOpen
  \bibfield  {author} {\bibinfo {author} {\bibfnamefont {P.~B.}\ \bibnamefont
  {Corkum}},\ }\enquote {\bibinfo {title} {Plasma perspective on strong field
  multiphoton ionization},}\ \href {\doibase 10.1103/PhysRevLett.71.1994}
  {\bibfield  {journal} {\bibinfo  {journal} {Phys. Rev. Lett.}\ }\textbf
  {\bibinfo {volume} {71}},\ \bibinfo {pages} {1994} (\bibinfo {year}
  {1993})}\BibitemShut {NoStop}%
\bibitem [{\citenamefont {Vampa}\ \emph {et~al.}(2014)\citenamefont {Vampa},
  \citenamefont {McDonald}, \citenamefont {Orlando}, \citenamefont {Klug},
  \citenamefont {Corkum},\ and\ \citenamefont {Brabec}}]{vampa2014prl}%
  \BibitemOpen
  \bibfield  {author} {\bibinfo {author} {\bibfnamefont {G.}~\bibnamefont
  {Vampa}}, \bibinfo {author} {\bibfnamefont {C.~R.}\ \bibnamefont {McDonald}},
  \bibinfo {author} {\bibfnamefont {G.}~\bibnamefont {Orlando}}, \bibinfo
  {author} {\bibfnamefont {D.~D.}\ \bibnamefont {Klug}}, \bibinfo {author}
  {\bibfnamefont {P.~B.}\ \bibnamefont {Corkum}}, \ and\ \bibinfo {author}
  {\bibfnamefont {T.}~\bibnamefont {Brabec}},\ }\enquote {\bibinfo {title}
  {Theoretical Analysis of High-Harmonic Generation in Solids},}\ \href
  {\doibase 10.1103/PhysRevLett.113.073901} {\bibfield  {journal} {\bibinfo
  {journal} {Phys. Rev. Lett.}\ }\textbf {\bibinfo {volume} {113}},\ \bibinfo
  {pages} {073901} (\bibinfo {year} {2014})}\BibitemShut {NoStop}%
\bibitem [{\citenamefont {Vampa}\ and\ \citenamefont
  {Brabec}(2017)}]{vampa2017jpb}%
  \BibitemOpen
  \bibfield  {author} {\bibinfo {author} {\bibfnamefont {G.}~\bibnamefont
  {Vampa}}\ and\ \bibinfo {author} {\bibfnamefont {T.}~\bibnamefont {Brabec}},\
  }\enquote {\bibinfo {title} {Merge of high harmonic generation from gases and
  solids and its implications for attosecond science},}\ \href {\doibase
  10.1088/1361-6455/aa528d} {\bibfield  {journal} {\bibinfo  {journal} {J.
  Phys. B}\ }\textbf {\bibinfo {volume} {50}},\ \bibinfo {pages} {083001}
  (\bibinfo {year} {2017})}\BibitemShut {NoStop}%
\bibitem [{\citenamefont {Higuchi}\ \emph {et~al.}(2014)\citenamefont
  {Higuchi}, \citenamefont {Stockman},\ and\ \citenamefont
  {Hommelhoff}}]{higuchi2014prl}%
  \BibitemOpen
  \bibfield  {author} {\bibinfo {author} {\bibfnamefont {T.}~\bibnamefont
  {Higuchi}}, \bibinfo {author} {\bibfnamefont {M.~I.}\ \bibnamefont
  {Stockman}}, \ and\ \bibinfo {author} {\bibfnamefont {P.}~\bibnamefont
  {Hommelhoff}},\ }\enquote {\bibinfo {title} {Strong-Field Perspective on
  High-Harmonic Radiation from Bulk Solids},}\ \href {\doibase
  10.1103/PhysRevLett.113.213901} {\bibfield  {journal} {\bibinfo  {journal}
  {Phys. Rev. Lett.}\ }\textbf {\bibinfo {volume} {113}},\ \bibinfo {pages}
  {213901} (\bibinfo {year} {2014})}\BibitemShut {NoStop}%
\bibitem [{\citenamefont {Vampa}\ \emph
  {et~al.}(2015{\natexlab{b}})\citenamefont {Vampa}, \citenamefont {McDonald},
  \citenamefont {Orlando}, \citenamefont {Corkum},\ and\ \citenamefont
  {Brabec}}]{vampa2015prb}%
  \BibitemOpen
  \bibfield  {author} {\bibinfo {author} {\bibfnamefont {G.}~\bibnamefont
  {Vampa}}, \bibinfo {author} {\bibfnamefont {C.~R.}\ \bibnamefont {McDonald}},
  \bibinfo {author} {\bibfnamefont {G.}~\bibnamefont {Orlando}}, \bibinfo
  {author} {\bibfnamefont {P.~B.}\ \bibnamefont {Corkum}}, \ and\ \bibinfo
  {author} {\bibfnamefont {T.}~\bibnamefont {Brabec}},\ }\enquote {\bibinfo
  {title} {Semiclassical analysis of high harmonic generation in bulk
  crystals},}\ \href {\doibase 10.1103/PhysRevB.91.064302} {\bibfield
  {journal} {\bibinfo  {journal} {Phys. Rev. B}\ }\textbf {\bibinfo {volume}
  {91}},\ \bibinfo {pages} {064302} (\bibinfo {year}
  {2015}{\natexlab{b}})}\BibitemShut {NoStop}%
\bibitem [{\citenamefont {McDonald}\ \emph {et~al.}(2015)\citenamefont
  {McDonald}, \citenamefont {Vampa}, \citenamefont {Corkum},\ and\
  \citenamefont {Brabec}}]{cmcdo2015pra}%
  \BibitemOpen
  \bibfield  {author} {\bibinfo {author} {\bibfnamefont {C.~R.}\ \bibnamefont
  {McDonald}}, \bibinfo {author} {\bibfnamefont {G.}~\bibnamefont {Vampa}},
  \bibinfo {author} {\bibfnamefont {P.~B.}\ \bibnamefont {Corkum}}, \ and\
  \bibinfo {author} {\bibfnamefont {T.}~\bibnamefont {Brabec}},\ }\enquote
  {\bibinfo {title} {Interband Bloch oscillation mechanism for high-harmonic
  generation in semiconductor crystals},}\ \href {\doibase
  10.1103/PhysRevA.92.033845} {\bibfield  {journal} {\bibinfo  {journal} {Phys.
  Rev. A}\ }\textbf {\bibinfo {volume} {92}},\ \bibinfo {pages} {033845}
  (\bibinfo {year} {2015})}\BibitemShut {NoStop}%
\bibitem [{\citenamefont {Hawkins}\ \emph {et~al.}(2015)\citenamefont
  {Hawkins}, \citenamefont {Ivanov},\ and\ \citenamefont
  {Yakovlev}}]{hawkins2015pra}%
  \BibitemOpen
  \bibfield  {author} {\bibinfo {author} {\bibfnamefont {P.~G.}\ \bibnamefont
  {Hawkins}}, \bibinfo {author} {\bibfnamefont {M.~Y.}\ \bibnamefont {Ivanov}},
  \ and\ \bibinfo {author} {\bibfnamefont {V.~S.}\ \bibnamefont {Yakovlev}},\
  }\enquote {\bibinfo {title} {Effect of multiple conduction bands on
  high-harmonic emission from dielectrics},}\ \href {\doibase
  10.1103/PhysRevA.91.013405} {\bibfield  {journal} {\bibinfo  {journal} {Phys.
  Rev. A}\ }\textbf {\bibinfo {volume} {91}},\ \bibinfo {pages} {013405}
  (\bibinfo {year} {2015})}\BibitemShut {NoStop}%
\bibitem [{\citenamefont {Wu}\ \emph {et~al.}(2015)\citenamefont {Wu},
  \citenamefont {Ghimire}, \citenamefont {Reis}, \citenamefont {Schafer},\ and\
  \citenamefont {Gaarde}}]{wmx2015pra}%
  \BibitemOpen
  \bibfield  {author} {\bibinfo {author} {\bibfnamefont {M.}~\bibnamefont
  {Wu}}, \bibinfo {author} {\bibfnamefont {S.}~\bibnamefont {Ghimire}},
  \bibinfo {author} {\bibfnamefont {D.~A.}\ \bibnamefont {Reis}}, \bibinfo
  {author} {\bibfnamefont {K.~J.}\ \bibnamefont {Schafer}}, \ and\ \bibinfo
  {author} {\bibfnamefont {M.~B.}\ \bibnamefont {Gaarde}},\ }\enquote {\bibinfo
  {title} {High-harmonic generation from Bloch electrons in solids},}\ \href
  {\doibase 10.1103/PhysRevA.91.043839} {\bibfield  {journal} {\bibinfo
  {journal} {Phys. Rev. A}\ }\textbf {\bibinfo {volume} {91}},\ \bibinfo
  {pages} {043839} (\bibinfo {year} {2015})}\BibitemShut {NoStop}%
\bibitem [{\citenamefont {Wu}\ \emph {et~al.}(2016)\citenamefont {Wu},
  \citenamefont {Browne}, \citenamefont {Schafer},\ and\ \citenamefont
  {Gaarde}}]{wmx2016pra}%
  \BibitemOpen
  \bibfield  {author} {\bibinfo {author} {\bibfnamefont {M.}~\bibnamefont
  {Wu}}, \bibinfo {author} {\bibfnamefont {D.~A.}\ \bibnamefont {Browne}},
  \bibinfo {author} {\bibfnamefont {K.~J.}\ \bibnamefont {Schafer}}, \ and\
  \bibinfo {author} {\bibfnamefont {M.~B.}\ \bibnamefont {Gaarde}},\ }\enquote
  {\bibinfo {title} {Multilevel perspective on high-order harmonic generation
  in solids},}\ \href {\doibase 10.1103/PhysRevA.94.063403} {\bibfield
  {journal} {\bibinfo  {journal} {Phys. Rev. A}\ }\textbf {\bibinfo {volume}
  {94}},\ \bibinfo {pages} {063403} (\bibinfo {year} {2016})}\BibitemShut
  {NoStop}%
\bibitem [{\citenamefont {Guan}\ \emph {et~al.}(2016)\citenamefont {Guan},
  \citenamefont {Zhou},\ and\ \citenamefont {Bian}}]{bxb2016pra}%
  \BibitemOpen
  \bibfield  {author} {\bibinfo {author} {\bibfnamefont {Z.}~\bibnamefont
  {Guan}}, \bibinfo {author} {\bibfnamefont {X.-X.}\ \bibnamefont {Zhou}}, \
  and\ \bibinfo {author} {\bibfnamefont {X.-B.}\ \bibnamefont {Bian}},\
  }\enquote {\bibinfo {title} {High-order-harmonic generation from periodic
  potentials driven by few-cycle laser pulses},}\ \href {\doibase
  10.1103/PhysRevA.93.033852} {\bibfield  {journal} {\bibinfo  {journal} {Phys.
  Rev. A}\ }\textbf {\bibinfo {volume} {93}},\ \bibinfo {pages} {033852}
  (\bibinfo {year} {2016})}\BibitemShut {NoStop}%
\bibitem [{\citenamefont {Du}\ \emph {et~al.}(2018{\natexlab{a}})\citenamefont
  {Du}, \citenamefont {Huang},\ and\ \citenamefont {Bian}}]{bxb2018pra1}%
  \BibitemOpen
  \bibfield  {author} {\bibinfo {author} {\bibfnamefont {T.-Y.}\ \bibnamefont
  {Du}}, \bibinfo {author} {\bibfnamefont {X.-H.}\ \bibnamefont {Huang}}, \
  and\ \bibinfo {author} {\bibfnamefont {X.-B.}\ \bibnamefont {Bian}},\
  }\enquote {\bibinfo {title} {High-order-harmonic generation from solids: The
  contributions of the Bloch wave packets moving at the group and phase
  velocities},}\ \href {\doibase 10.1103/PhysRevA.97.013403} {\bibfield
  {journal} {\bibinfo  {journal} {Phys. Rev. A}\ }\textbf {\bibinfo {volume}
  {97}},\ \bibinfo {pages} {013403} (\bibinfo {year}
  {2018}{\natexlab{a}})}\BibitemShut {NoStop}%
\bibitem [{\citenamefont {Du}\ \emph {et~al.}(2018{\natexlab{b}})\citenamefont
  {Du}, \citenamefont {Tang}, \citenamefont {Huang},\ and\ \citenamefont
  {Bian}}]{bxb2018pra2}%
  \BibitemOpen
  \bibfield  {author} {\bibinfo {author} {\bibfnamefont {T.-Y.}\ \bibnamefont
  {Du}}, \bibinfo {author} {\bibfnamefont {D.}~\bibnamefont {Tang}}, \bibinfo
  {author} {\bibfnamefont {X.-H.}\ \bibnamefont {Huang}}, \ and\ \bibinfo
  {author} {\bibfnamefont {X.-B.}\ \bibnamefont {Bian}},\ }\enquote {\bibinfo
  {title} {Multichannel high-order harmonic generation from solids},}\ \href
  {\doibase 10.1103/PhysRevA.97.043413} {\bibfield  {journal} {\bibinfo
  {journal} {Phys. Rev. A}\ }\textbf {\bibinfo {volume} {97}},\ \bibinfo
  {pages} {043413} (\bibinfo {year} {2018}{\natexlab{b}})}\BibitemShut
  {NoStop}%
\bibitem [{\citenamefont {Jin}\ \emph {et~al.}(2018)\citenamefont {Jin},
  \citenamefont {Liang}, \citenamefont {Xiao}, \citenamefont {Wang},
  \citenamefont {Chen}, \citenamefont {Wu}, \citenamefont {Gong},\ and\
  \citenamefont {Peng}}]{jin2018jpb}%
  \BibitemOpen
  \bibfield  {author} {\bibinfo {author} {\bibfnamefont {J.-Z.}\ \bibnamefont
  {Jin}}, \bibinfo {author} {\bibfnamefont {H.}~\bibnamefont {Liang}}, \bibinfo
  {author} {\bibfnamefont {X.-R.}\ \bibnamefont {Xiao}}, \bibinfo {author}
  {\bibfnamefont {M.-X.}\ \bibnamefont {Wang}}, \bibinfo {author}
  {\bibfnamefont {S.-G.}\ \bibnamefont {Chen}}, \bibinfo {author}
  {\bibfnamefont {X.-Y.}\ \bibnamefont {Wu}}, \bibinfo {author} {\bibfnamefont
  {Q.}~\bibnamefont {Gong}}, \ and\ \bibinfo {author} {\bibfnamefont {L.-Y.}\
  \bibnamefont {Peng}},\ }\enquote {\bibinfo {title} {Michelson interferometry
  of high-order harmonic generation in solids},}\ \href {\doibase
  10.1088/1361-6455/aad40b} {\bibfield  {journal} {\bibinfo  {journal} {J.
  Phys. B}\ }\textbf {\bibinfo {volume} {51}},\ \bibinfo {pages} {16LT01}
  (\bibinfo {year} {2018})}\BibitemShut {NoStop}%
\bibitem [{\citenamefont {Luu}\ and\ \citenamefont
  {W\"orner}(2016)}]{luu2016prb}%
  \BibitemOpen
  \bibfield  {author} {\bibinfo {author} {\bibfnamefont {T.~T.}\ \bibnamefont
  {Luu}}\ and\ \bibinfo {author} {\bibfnamefont {H.~J.}\ \bibnamefont
  {W\"orner}},\ }\enquote {\bibinfo {title} {High-order harmonic generation in
  solids: A unifying approach},}\ \href {\doibase 10.1103/PhysRevB.94.115164}
  {\bibfield  {journal} {\bibinfo  {journal} {Phys. Rev. B}\ }\textbf {\bibinfo
  {volume} {94}},\ \bibinfo {pages} {115164} (\bibinfo {year}
  {2016})}\BibitemShut {NoStop}%
\bibitem [{\citenamefont {F\"oldi}(2017)}]{foeldi2017prb}%
  \BibitemOpen
  \bibfield  {author} {\bibinfo {author} {\bibfnamefont {P.}~\bibnamefont
  {F\"oldi}},\ }\enquote {\bibinfo {title} {Gauge invariance and interpretation
  of interband and intraband processes in high-order harmonic generation from
  bulk solids},}\ \href {\doibase 10.1103/PhysRevB.96.035112} {\bibfield
  {journal} {\bibinfo  {journal} {Phys. Rev. B}\ }\textbf {\bibinfo {volume}
  {96}},\ \bibinfo {pages} {035112} (\bibinfo {year} {2017})}\BibitemShut
  {NoStop}%
\bibitem [{\citenamefont {Osika}\ \emph {et~al.}(2017)\citenamefont {Osika},
  \citenamefont {Chac\'on}, \citenamefont {Ortmann}, \citenamefont {Su\'arez},
  \citenamefont {P\'erez-Hern\'andez}, \citenamefont {Szafran}, \citenamefont
  {Ciappina}, \citenamefont {Sols}, \citenamefont {Landsman},\ and\
  \citenamefont {Lewenstein}}]{osika2017prx}%
  \BibitemOpen
  \bibfield  {author} {\bibinfo {author} {\bibfnamefont {E.~N.}\ \bibnamefont
  {Osika}}, \bibinfo {author} {\bibfnamefont {A.}~\bibnamefont {Chac\'on}},
  \bibinfo {author} {\bibfnamefont {L.}~\bibnamefont {Ortmann}}, \bibinfo
  {author} {\bibfnamefont {N.}~\bibnamefont {Su\'arez}}, \bibinfo {author}
  {\bibfnamefont {J.~A.}\ \bibnamefont {P\'erez-Hern\'andez}}, \bibinfo
  {author} {\bibfnamefont {B.}~\bibnamefont {Szafran}}, \bibinfo {author}
  {\bibfnamefont {M.~F.}\ \bibnamefont {Ciappina}}, \bibinfo {author}
  {\bibfnamefont {F.}~\bibnamefont {Sols}}, \bibinfo {author} {\bibfnamefont
  {A.~S.}\ \bibnamefont {Landsman}}, \ and\ \bibinfo {author} {\bibfnamefont
  {M.}~\bibnamefont {Lewenstein}},\ }\enquote {\bibinfo {title} {Wannier-Bloch
  Approach to Localization in High-Harmonics Generation in Solids},}\ \href
  {\doibase 10.1103/PhysRevX.7.021017} {\bibfield  {journal} {\bibinfo
  {journal} {Phys. Rev. X}\ }\textbf {\bibinfo {volume} {7}},\ \bibinfo {pages}
  {021017} (\bibinfo {year} {2017})}\BibitemShut {NoStop}%
\bibitem [{\citenamefont {Tancogne-Dejean}\ \emph {et~al.}(2017)\citenamefont
  {Tancogne-Dejean}, \citenamefont {M\"ucke}, \citenamefont {K\"artner},\ and\
  \citenamefont {Rubio}}]{rubio2017prl}%
  \BibitemOpen
  \bibfield  {author} {\bibinfo {author} {\bibfnamefont {N.}~\bibnamefont
  {Tancogne-Dejean}}, \bibinfo {author} {\bibfnamefont {O.~D.}\ \bibnamefont
  {M\"ucke}}, \bibinfo {author} {\bibfnamefont {F.~X.}\ \bibnamefont
  {K\"artner}}, \ and\ \bibinfo {author} {\bibfnamefont {A.}~\bibnamefont
  {Rubio}},\ }\enquote {\bibinfo {title} {Impact of the Electronic Band
  Structure in High-Harmonic Generation Spectra of Solids},}\ \href {\doibase
  10.1103/PhysRevLett.118.087403} {\bibfield  {journal} {\bibinfo  {journal}
  {Phys. Rev. Lett.}\ }\textbf {\bibinfo {volume} {118}},\ \bibinfo {pages}
  {087403} (\bibinfo {year} {2017})}\BibitemShut {NoStop}%
\bibitem [{\citenamefont {Hansen}\ \emph {et~al.}(2017)\citenamefont {Hansen},
  \citenamefont {Deffge},\ and\ \citenamefont {Bauer}}]{kkh2017pra}%
  \BibitemOpen
  \bibfield  {author} {\bibinfo {author} {\bibfnamefont {K.~K.}\ \bibnamefont
  {Hansen}}, \bibinfo {author} {\bibfnamefont {T.}~\bibnamefont {Deffge}}, \
  and\ \bibinfo {author} {\bibfnamefont {D.}~\bibnamefont {Bauer}},\ }\enquote
  {\bibinfo {title} {High-order harmonic generation in solid slabs beyond the
  single-active-electron approximation},}\ \href {\doibase
  10.1103/PhysRevA.96.053418} {\bibfield  {journal} {\bibinfo  {journal} {Phys.
  Rev. A}\ }\textbf {\bibinfo {volume} {96}},\ \bibinfo {pages} {053418}
  (\bibinfo {year} {2017})}\BibitemShut {NoStop}%
\bibitem [{\citenamefont {Hansen}\ \emph {et~al.}(2018)\citenamefont {Hansen},
  \citenamefont {Bauer},\ and\ \citenamefont {Madsen}}]{kkh2018pra}%
  \BibitemOpen
  \bibfield  {author} {\bibinfo {author} {\bibfnamefont {K.~K.}\ \bibnamefont
  {Hansen}}, \bibinfo {author} {\bibfnamefont {D.}~\bibnamefont {Bauer}}, \
  and\ \bibinfo {author} {\bibfnamefont {L.~B.}\ \bibnamefont {Madsen}},\
  }\enquote {\bibinfo {title} {Finite-system effects on high-order harmonic
  generation: From atoms to solids},}\ \href {\doibase
  10.1103/PhysRevA.97.043424} {\bibfield  {journal} {\bibinfo  {journal} {Phys.
  Rev. A}\ }\textbf {\bibinfo {volume} {97}},\ \bibinfo {pages} {043424}
  (\bibinfo {year} {2018})}\BibitemShut {NoStop}%
\bibitem [{\citenamefont {Bauer}\ and\ \citenamefont
  {Hansen}(2018)}]{kkh2018prl}%
  \BibitemOpen
  \bibfield  {author} {\bibinfo {author} {\bibfnamefont {D.}~\bibnamefont
  {Bauer}}\ and\ \bibinfo {author} {\bibfnamefont {K.~K.}\ \bibnamefont
  {Hansen}},\ }\enquote {\bibinfo {title} {High-Harmonic Generation in Solids
  with and without Topological Edge States},}\ \href {\doibase
  10.1103/PhysRevLett.120.177401} {\bibfield  {journal} {\bibinfo  {journal}
  {Phys. Rev. Lett.}\ }\textbf {\bibinfo {volume} {120}},\ \bibinfo {pages}
  {177401} (\bibinfo {year} {2018})}\BibitemShut {NoStop}%
\bibitem [{\citenamefont {Murakami}\ \emph {et~al.}(2018)\citenamefont
  {Murakami}, \citenamefont {Eckstein},\ and\ \citenamefont
  {Werner}}]{murakami2018prl}%
  \BibitemOpen
  \bibfield  {author} {\bibinfo {author} {\bibfnamefont {Y.}~\bibnamefont
  {Murakami}}, \bibinfo {author} {\bibfnamefont {M.}~\bibnamefont {Eckstein}},
  \ and\ \bibinfo {author} {\bibfnamefont {P.}~\bibnamefont {Werner}},\
  }\enquote {\bibinfo {title} {High-Harmonic Generation in Mott Insulators},}\
  \href {\doibase 10.1103/PhysRevLett.121.057405} {\bibfield  {journal}
  {\bibinfo  {journal} {Phys. Rev. Lett.}\ }\textbf {\bibinfo {volume} {121}},\
  \bibinfo {pages} {057405} (\bibinfo {year} {2018})}\BibitemShut {NoStop}%
\bibitem [{\citenamefont {Silva}\ \emph {et~al.}(2018)\citenamefont {Silva},
  \citenamefont {Blinov}, \citenamefont {Rubtsov}, \citenamefont {Smirnova},\
  and\ \citenamefont {Ivanov}}]{silva2018natphoton}%
  \BibitemOpen
  \bibfield  {author} {\bibinfo {author} {\bibfnamefont {R.}~\bibnamefont
  {Silva}}, \bibinfo {author} {\bibfnamefont {I.~V.}\ \bibnamefont {Blinov}},
  \bibinfo {author} {\bibfnamefont {A.~N.}\ \bibnamefont {Rubtsov}}, \bibinfo
  {author} {\bibfnamefont {O.}~\bibnamefont {Smirnova}}, \ and\ \bibinfo
  {author} {\bibfnamefont {M.}~\bibnamefont {Ivanov}},\ }\enquote {\bibinfo
  {title} {High-harmonic spectroscopy of ultrafast many-body dynamics in
  strongly correlated systems},}\ \href {\doibase 10.1038/s41566-018-0129-0}
  {\bibfield  {journal} {\bibinfo  {journal} {Nat. Photon.}\ }\textbf {\bibinfo
  {volume} {12}},\ \bibinfo {pages} {266} (\bibinfo {year} {2018})}\BibitemShut
  {NoStop}%
\bibitem [{\citenamefont {Luu}\ and\ \citenamefont
  {W{\"o}rner}(2018)}]{luu2018natcommun}%
  \BibitemOpen
  \bibfield  {author} {\bibinfo {author} {\bibfnamefont {T.~T.}\ \bibnamefont
  {Luu}}\ and\ \bibinfo {author} {\bibfnamefont {H.~J.}\ \bibnamefont
  {W{\"o}rner}},\ }\enquote {\bibinfo {title} {Measurement of the Berry
  curvature of solids using high-harmonic spectroscopy},}\ \href {\doibase
  10.1038/s41467-018-03397-4} {\bibfield  {journal} {\bibinfo  {journal} {Nat.
  Commun.}\ }\textbf {\bibinfo {volume} {9}},\ \bibinfo {pages} {916} (\bibinfo
  {year} {2018})}\BibitemShut {NoStop}%
\bibitem [{\citenamefont {Sivis}\ \emph {et~al.}(2017)\citenamefont {Sivis},
  \citenamefont {Taucer}, \citenamefont {Vampa}, \citenamefont {Johnston},
  \citenamefont {Staudte}, \citenamefont {Naumov}, \citenamefont {Villeneuve},
  \citenamefont {Ropers},\ and\ \citenamefont {Corkum}}]{sivis2017science}%
  \BibitemOpen
  \bibfield  {author} {\bibinfo {author} {\bibfnamefont {M.}~\bibnamefont
  {Sivis}}, \bibinfo {author} {\bibfnamefont {M.}~\bibnamefont {Taucer}},
  \bibinfo {author} {\bibfnamefont {G.}~\bibnamefont {Vampa}}, \bibinfo
  {author} {\bibfnamefont {K.}~\bibnamefont {Johnston}}, \bibinfo {author}
  {\bibfnamefont {A.}~\bibnamefont {Staudte}}, \bibinfo {author} {\bibfnamefont
  {A.~Y.}\ \bibnamefont {Naumov}}, \bibinfo {author} {\bibfnamefont {D.~M.}\
  \bibnamefont {Villeneuve}}, \bibinfo {author} {\bibfnamefont
  {C.}~\bibnamefont {Ropers}}, \ and\ \bibinfo {author} {\bibfnamefont {P.~B.}\
  \bibnamefont {Corkum}},\ }\enquote {\bibinfo {title} {Tailored semiconductors
  for high-harmonic optoelectronics},}\ \href {\doibase
  10.1126/science.aan2395} {\bibfield  {journal} {\bibinfo  {journal}
  {Science}\ }\textbf {\bibinfo {volume} {357}},\ \bibinfo {pages} {303}
  (\bibinfo {year} {2017})}\BibitemShut {NoStop}%
\bibitem [{\citenamefont {McDonald}\ \emph {et~al.}(2017)\citenamefont
  {McDonald}, \citenamefont {Amin}, \citenamefont {Aalmalki},\ and\
  \citenamefont {Brabec}}]{cmcdo2017prl2}%
  \BibitemOpen
  \bibfield  {author} {\bibinfo {author} {\bibfnamefont {C.~R.}\ \bibnamefont
  {McDonald}}, \bibinfo {author} {\bibfnamefont {K.~S.}\ \bibnamefont {Amin}},
  \bibinfo {author} {\bibfnamefont {S.}~\bibnamefont {Aalmalki}}, \ and\
  \bibinfo {author} {\bibfnamefont {T.}~\bibnamefont {Brabec}},\ }\enquote
  {\bibinfo {title} {Enhancing High Harmonic Output in Solids through Quantum
  Confinement},}\ \href {\doibase 10.1103/PhysRevLett.119.183902} {\bibfield
  {journal} {\bibinfo  {journal} {Phys. Rev. Lett.}\ }\textbf {\bibinfo
  {volume} {119}},\ \bibinfo {pages} {183902} (\bibinfo {year}
  {2017})}\BibitemShut {NoStop}%
\bibitem [{\citenamefont {Du}\ \emph {et~al.}(2016)\citenamefont {Du},
  \citenamefont {Guan}, \citenamefont {Zhou},\ and\ \citenamefont
  {Bian}}]{bxb2016pra2}%
  \BibitemOpen
  \bibfield  {author} {\bibinfo {author} {\bibfnamefont {T.-Y.}\ \bibnamefont
  {Du}}, \bibinfo {author} {\bibfnamefont {Z.}~\bibnamefont {Guan}}, \bibinfo
  {author} {\bibfnamefont {X.-X.}\ \bibnamefont {Zhou}}, \ and\ \bibinfo
  {author} {\bibfnamefont {X.-B.}\ \bibnamefont {Bian}},\ }\enquote {\bibinfo
  {title} {Enhanced high-order harmonic generation from periodic potentials in
  inhomogeneous laser fields},}\ \href {\doibase 10.1103/PhysRevA.94.023419}
  {\bibfield  {journal} {\bibinfo  {journal} {Phys. Rev. A}\ }\textbf {\bibinfo
  {volume} {94}},\ \bibinfo {pages} {023419} (\bibinfo {year}
  {2016})}\BibitemShut {NoStop}%
\bibitem [{\citenamefont {Huang}\ \emph {et~al.}(2017)\citenamefont {Huang},
  \citenamefont {Zhu}, \citenamefont {Li}, \citenamefont {Liu}, \citenamefont
  {Lan},\ and\ \citenamefont {Lu}}]{huang2017pra}%
  \BibitemOpen
  \bibfield  {author} {\bibinfo {author} {\bibfnamefont {T.}~\bibnamefont
  {Huang}}, \bibinfo {author} {\bibfnamefont {X.}~\bibnamefont {Zhu}}, \bibinfo
  {author} {\bibfnamefont {L.}~\bibnamefont {Li}}, \bibinfo {author}
  {\bibfnamefont {X.}~\bibnamefont {Liu}}, \bibinfo {author} {\bibfnamefont
  {P.}~\bibnamefont {Lan}}, \ and\ \bibinfo {author} {\bibfnamefont
  {P.}~\bibnamefont {Lu}},\ }\enquote {\bibinfo {title} {High-order-harmonic
  generation of a doped semiconductor},}\ \href {\doibase
  10.1103/PhysRevA.96.043425} {\bibfield  {journal} {\bibinfo  {journal} {Phys.
  Rev. A}\ }\textbf {\bibinfo {volume} {96}},\ \bibinfo {pages} {043425}
  (\bibinfo {year} {2017})}\BibitemShut {NoStop}%
\bibitem [{\citenamefont {Burgess}\ \emph {et~al.}(2016)\citenamefont
  {Burgess}, \citenamefont {Saxena}, \citenamefont {Mokkapati}, \citenamefont
  {Li}, \citenamefont {Hall}, \citenamefont {Davis}, \citenamefont {Wang},
  \citenamefont {Smith}, \citenamefont {Fu}, \citenamefont {Caroff} \emph
  {et~al.}}]{impurity2016natcommun}%
  \BibitemOpen
  \bibfield  {author} {\bibinfo {author} {\bibfnamefont {T.}~\bibnamefont
  {Burgess}}, \bibinfo {author} {\bibfnamefont {D.}~\bibnamefont {Saxena}},
  \bibinfo {author} {\bibfnamefont {S.}~\bibnamefont {Mokkapati}}, \bibinfo
  {author} {\bibfnamefont {Z.}~\bibnamefont {Li}}, \bibinfo {author}
  {\bibfnamefont {C.~R.}\ \bibnamefont {Hall}}, \bibinfo {author}
  {\bibfnamefont {J.~A.}\ \bibnamefont {Davis}}, \bibinfo {author}
  {\bibfnamefont {Y.}~\bibnamefont {Wang}}, \bibinfo {author} {\bibfnamefont
  {L.~M.}\ \bibnamefont {Smith}}, \bibinfo {author} {\bibfnamefont
  {L.}~\bibnamefont {Fu}}, \bibinfo {author} {\bibfnamefont {P.}~\bibnamefont
  {Caroff}},  \emph {et~al.},\ }\enquote {\bibinfo {title} {Doping-enhanced
  radiative efficiency enables lasing in unpassivated GaAs nanowires},}\ \href
  {\doibase 10.1038/ncomms11927} {\bibfield  {journal} {\bibinfo  {journal}
  {Nat. Commun.}\ }\textbf {\bibinfo {volume} {7}},\ \bibinfo {pages} {11927}
  (\bibinfo {year} {2016})}\BibitemShut {NoStop}%
\bibitem [{\citenamefont {Klyukin}\ \emph {et~al.}(2018)\citenamefont
  {Klyukin}, \citenamefont {Tao}, \citenamefont {Tsymbal},\ and\ \citenamefont
  {Alexandrov}}]{impurity2018prl}%
  \BibitemOpen
  \bibfield  {author} {\bibinfo {author} {\bibfnamefont {K.}~\bibnamefont
  {Klyukin}}, \bibinfo {author} {\bibfnamefont {L.~L.}\ \bibnamefont {Tao}},
  \bibinfo {author} {\bibfnamefont {E.~Y.}\ \bibnamefont {Tsymbal}}, \ and\
  \bibinfo {author} {\bibfnamefont {V.}~\bibnamefont {Alexandrov}},\ }\enquote
  {\bibinfo {title} {Defect-Assisted Tunneling Electroresistance in
  Ferroelectric Tunnel Junctions},}\ \href {\doibase
  10.1103/PhysRevLett.121.056601} {\bibfield  {journal} {\bibinfo  {journal}
  {Phys. Rev. Lett.}\ }\textbf {\bibinfo {volume} {121}},\ \bibinfo {pages}
  {056601} (\bibinfo {year} {2018})}\BibitemShut {NoStop}%
\bibitem [{Note1()}]{Note1}%
  \BibitemOpen
  \bibinfo {note} {The periodic arrangement of dopants at a very high impurity
  concentration considered in the single-active-electron model calculations of
  Ref.~\cite {huang2017pra} is atypical and leads to a substantial change of
  the system.}\BibitemShut {Stop}%
\bibitem [{\citenamefont {Runge}\ and\ \citenamefont
  {Gross}(1984)}]{tddft1984prl}%
  \BibitemOpen
  \bibfield  {author} {\bibinfo {author} {\bibfnamefont {E.}~\bibnamefont
  {Runge}}\ and\ \bibinfo {author} {\bibfnamefont {E.~K.~U.}\ \bibnamefont
  {Gross}},\ }\enquote {\bibinfo {title} {Density-Functional Theory for
  Time-Dependent Systems},}\ \href {\doibase 10.1103/PhysRevLett.52.997}
  {\bibfield  {journal} {\bibinfo  {journal} {Phys. Rev. Lett.}\ }\textbf
  {\bibinfo {volume} {52}},\ \bibinfo {pages} {997} (\bibinfo {year}
  {1984})}\BibitemShut {NoStop}%
\bibitem [{\citenamefont {Ullrich}(2011)}]{ullrich2011book}%
  \BibitemOpen
  \bibfield  {author} {\bibinfo {author} {\bibfnamefont {C.~A.}\ \bibnamefont
  {Ullrich}},\ }\href@noop {} {\emph {\bibinfo {title} {Time-Dependent
  Density-Functional Theory: Concepts and Applications}}}\ (\bibinfo
  {publisher} {Oxford University Press, Oxford},\ \bibinfo {year}
  {2011})\BibitemShut {NoStop}%
\bibitem [{\citenamefont {Pantelides}(1978)}]{impurity1978rmp}%
  \BibitemOpen
  \bibfield  {author} {\bibinfo {author} {\bibfnamefont {S.~T.}\ \bibnamefont
  {Pantelides}},\ }\enquote {\bibinfo {title} {The electronic structure of
  impurities and other point defects in semiconductors},}\ \href {\doibase
  10.1103/RevModPhys.50.797} {\bibfield  {journal} {\bibinfo  {journal} {Rev.
  Mod. Phys.}\ }\textbf {\bibinfo {volume} {50}},\ \bibinfo {pages} {797}
  (\bibinfo {year} {1978})}\BibitemShut {NoStop}%
\bibitem [{\citenamefont {Bauer}(2017)}]{bauer2017book}%
  \BibitemOpen
  \bibinfo {editor} {\bibfnamefont {D.}~\bibnamefont {Bauer}},\ ed.,\
  \href@noop {} {\emph {\bibinfo {title} {Computational Strong-field Quantum
  Dynamics: Intense Light-matter Interactions}}}\ (\bibinfo  {publisher} {De
  Gruyter, Berlin},\ \bibinfo {year} {2017})\BibitemShut {NoStop}%
\bibitem [{\citenamefont {Slater}(1949)}]{slater1949pr}%
  \BibitemOpen
  \bibfield  {author} {\bibinfo {author} {\bibfnamefont {J.~C.}\ \bibnamefont
  {Slater}},\ }\enquote {\bibinfo {title} {Electrons in Perturbed Periodic
  Lattices},}\ \href {\doibase 10.1103/PhysRev.76.1592} {\bibfield  {journal}
  {\bibinfo  {journal} {Phys. Rev.}\ }\textbf {\bibinfo {volume} {76}},\
  \bibinfo {pages} {1592} (\bibinfo {year} {1949})}\BibitemShut {NoStop}%
\bibitem [{\citenamefont {Luttinger}\ and\ \citenamefont
  {Kohn}(1955)}]{luttinger1955pr}%
  \BibitemOpen
  \bibfield  {author} {\bibinfo {author} {\bibfnamefont {J.~M.}\ \bibnamefont
  {Luttinger}}\ and\ \bibinfo {author} {\bibfnamefont {W.}~\bibnamefont
  {Kohn}},\ }\enquote {\bibinfo {title} {Motion of Electrons and Holes in
  Perturbed Periodic Fields},}\ \href {\doibase 10.1103/PhysRev.97.869}
  {\bibfield  {journal} {\bibinfo  {journal} {Phys. Rev.}\ }\textbf {\bibinfo
  {volume} {97}},\ \bibinfo {pages} {869} (\bibinfo {year} {1955})}\BibitemShut
  {NoStop}%
\bibitem [{\citenamefont {Kohn}(1957)}]{impurity1957ssp}%
  \BibitemOpen
  \bibfield  {author} {\bibinfo {author} {\bibfnamefont {W.}~\bibnamefont
  {Kohn}},\ }\enquote {\bibinfo {title} {Shallow Impurity States in Silicon and
  Germanium},}\ \href {\doibase 10.1016/S0081-1947(08)60104-6} {\ \bibinfo
  {series} {Solid State Phys.},\ \textbf {\bibinfo {volume} {5}},\ \bibinfo
  {pages} {257} (\bibinfo {year} {1957})}\BibitemShut {NoStop}%
\bibitem [{\citenamefont {Bassani}\ \emph {et~al.}(1974)\citenamefont
  {Bassani}, \citenamefont {Iadonisi},\ and\ \citenamefont
  {Preziosi}}]{impurity1974rpp}%
  \BibitemOpen
  \bibfield  {author} {\bibinfo {author} {\bibfnamefont {F.}~\bibnamefont
  {Bassani}}, \bibinfo {author} {\bibfnamefont {G.}~\bibnamefont {Iadonisi}}, \
  and\ \bibinfo {author} {\bibfnamefont {B.}~\bibnamefont {Preziosi}},\
  }\enquote {\bibinfo {title} {Electronic impurity levels in semiconductors},}\
  \href {\doibase 10.1088/0034-4885/37/9/001} {\bibfield  {journal} {\bibinfo
  {journal} {Rep. Prog. Phys.}\ }\textbf {\bibinfo {volume} {37}},\ \bibinfo
  {pages} {1099} (\bibinfo {year} {1974})}\BibitemShut {NoStop}%
\bibitem [{\citenamefont {Keldysh}(1964)}]{keldysh1965jetp}%
  \BibitemOpen
  \bibfield  {author} {\bibinfo {author} {\bibfnamefont {L.~V.}\ \bibnamefont
  {Keldysh}},\ }\enquote {\bibinfo {title} {Ionization in the field of a strong
  electromagnetic wave},}\ \href@noop {} {\bibfield  {journal} {\bibinfo
  {journal} {Zh. Eksp. Teor. Fiz.}\ }\textbf {\bibinfo {volume} {47}},\
  \bibinfo {pages} {1945} (\bibinfo {year} {1964})},\ \bibinfo {note} {[Sov.
  Phys. JETP \textbf{20}, 1307 (1965)]}\BibitemShut {NoStop}%
\bibitem [{\citenamefont {Almalki}\ \emph {et~al.}(2018)\citenamefont
  {Almalki}, \citenamefont {Parks}, \citenamefont {Bart}, \citenamefont
  {Corkum}, \citenamefont {Brabec},\ and\ \citenamefont
  {McDonald}}]{impurity2018prb}%
  \BibitemOpen
  \bibfield  {author} {\bibinfo {author} {\bibfnamefont {S.}~\bibnamefont
  {Almalki}}, \bibinfo {author} {\bibfnamefont {A.~M.}\ \bibnamefont {Parks}},
  \bibinfo {author} {\bibfnamefont {G.}~\bibnamefont {Bart}}, \bibinfo {author}
  {\bibfnamefont {P.~B.}\ \bibnamefont {Corkum}}, \bibinfo {author}
  {\bibfnamefont {T.}~\bibnamefont {Brabec}}, \ and\ \bibinfo {author}
  {\bibfnamefont {C.~R.}\ \bibnamefont {McDonald}},\ }\enquote {\bibinfo
  {title} {High harmonic generation tomography of impurities in solids:
  Conceptual analysis},}\ \href {\doibase 10.1103/PhysRevB.98.144307}
  {\bibfield  {journal} {\bibinfo  {journal} {Phys. Rev. B}\ }\textbf {\bibinfo
  {volume} {98}},\ \bibinfo {pages} {144307} (\bibinfo {year}
  {2018})}\BibitemShut {NoStop}%
\bibitem [{Note2()}]{Note2}%
  \BibitemOpen
  \bibinfo {note} {For the considered vector-potential amplitude
  $A_{0}=0.22<\pi /(2a)$, we simply restrict to a single conduction band, CB1,
  since the semiclassical electron will not reach the first-Brillouin-zone
  boundary [Eq.~\protect \textup {\hbox {\mathsurround \z@ \protect \normalfont
  (\ignorespaces \ref {eq:cbmotion}\unskip \@@italiccorr )}}].}\BibitemShut
  {Stop}%
\end{thebibliography}

\end{document}